\documentclass[aps,prd,twocolumn,eqsecnum]{revtex4}

\def\bydefn{\stackrel{def}{=}}
\def\non{\nonumber}
\def\dprime{ {\prime \prime} }

\usepackage{amsfonts}
\usepackage{amsmath}
\usepackage{graphicx}
\usepackage{bm}
\usepackage{mathrsfs}
\begin{document}

\title{Self-Force with a Stochastic Component from Radiation
Reaction \\ of a Scalar Charge Moving in Curved Spacetime}
%Induced Fluctuations the : on Relativistic Particles in Gravitational Fields}
%\title{Stochastic Self-Force from Radiation Reaction of a Scalar Charge in Curved Spacetime}

\author{Chad R. Galley and B. L. Hu}
\affiliation{Department of Physics, University of Maryland,
College Park, Maryland, 20742}

\date{May 16, 2005}
\begin{abstract}
%Using the open quantum system and effective field theory paradigms
We give a quantum field theoretical derivation of  the scalar
Abraham-Lorentz-Dirac (ALD) equation and the self-force for a
scalar charged particle interacting with a quantum scalar field in
curved spacetime. We regularize the causal Green's function using a
quasi-local expansion in the spirit of effective field theory and
obtain a regular expression for the self-force. The scalar ALD
equation obtained in this way for the classical  motion of the
particle checks with the equation obtained by Quinn earlier
\cite{Quinn}.
%assuming that the particle's worldine is sufficiently decohered through interactions with the quantum field.
%Our approach requires us to extend Synge's world function formalism to non-geodesic particle motions, which we formulate in an Appendix.
We further derive a scalar ALD-Langevin equation with a classical
stochastic force accounting for the effect of quantum
fluctuations in the field,  which causes small fluctuations on the
particle trajectory. This equation will be useful for the study
of stochastic motion of charges under the influence of both
quantum and classical noise sources, derived either
self-consistently (as done here) or put in by hand (with
warnings). We show the possibility of secular effects from such
stochastic influences on the trajectory that may impact on the
present calculations of gravitational waveform templates.
\end{abstract}

\maketitle

\section{Introduction}

Interest in the problem of radiation reaction from particle
motion in a curved spacetime has seen a rapid increase in recent
years. The back-reaction of emitted radiation on the particle,
known as the self-force, changes the particle trajectories (e.g.,
near a black hole) from a simple geodesic motion. The
determination of the self force is essential to precision
calculations of particle trajectories and the determination of
waveforms from prospective astrophysical sources. Electromagnetic
radiation reaction in a curved spacetime was first studied by
DeWitt and Brehm \cite{DeWBre}. The gravitational radiation
reaction equation was first obtained by Mino, Sasaki and Tanaka
\cite{MST} and Quinn and Wald \cite{QW} and others, notably
Detweiler and Whiting \cite{Detweiler_Whiting}. The equation of
motion governing a scalar charge with radiation reaction was
first obtained by Quinn \cite{Quinn}. For an excellent review, see
\cite{Poisson}.

Parallel to this there has been detailed work devoted to particles
and detectors (e.g., atoms) moving in a quantum field and in
design studies of possible detection of Unruh radiation (see,
e.g.,\cite{HRCapri,LH1,RHA} and references therein). The
introduction of worldline path integral methods (see \cite{JH1}
and references therein) enable one to obtain equations of motion
for the charges and the field self-consistently. The introduction
of open system concepts and the influence functional method
enables one to derive stochastic equations with a noise source
derived {\it ab initio} and in a self consistent manner (See
\cite{RHA,RHK,RavalPhD} for accelerating detectors,
\cite{JH1,JHIARD,JohnsonPhD} for moving charges.)

In this paper, we use the worldline influence functional method
to study a particle with a scalar charge moving in its own
quantum scalar field in a curved spacetime. This is a
generalization of results obtained in \cite{JH1} to curved
spacetime. We are interested in the radiation it emits and its
backreaction (radiation reaction) on the trajectory of the
particle
%behavior of a charged particle as it emits scalar radiation in a (possibly) strong graviational field.
%Using the Feynman-Vernon influence functional we will describe the particle dynamics with effects from the quantum scalar field
and derive the equations of motion for the quantum average
(expectation value) of the particle's position. For those particle
trajectory histories which become sufficiently decohered, the
expectation value behaves classically. The scalar ALD equation we
derive in this limit checks with the result of Quinn \cite{Quinn}.
We then include in our consideration the effect of fluctuations
in the quantum field. We show how it behaves like a classical
stochastic force and derive a scalar ALD-Langevin equation for
the particle dynamics with a stochastic component, thus capturing
the induced small fluctuations on the particle trajectory. This
equation will be useful for the study of stochastic motion of
charges under the influence of both quantum or classical noise
sources, derived either self-consistently (as done here) or put
in by hand (with warnings). For astrophysical sources with some
stochastic component this effect may need to be included in more
accurate calculations of waveform templates.

In Sec. 2 we describe the worldline influence functional method
and how to obtain the semiclassical and stochastic particle
dynamics. In Sec. 3 we discuss how to regularize the causal Green
function in the spirit of effective field theory using a
quasi-local expansion. In Sec. 4 we derive the scalar ALD
equation. In Sec. 5 we derive the stochastic scalar ALD or the
scalar ALD-Langevin equation. We discuss the non-Markovian nature
of the noise induced effects and the possibility of secular
effects from such stochastic influences on the trajectory. In
Sec. 6 we discuss an array of issues pertinent to the present
problem and approach. In Sec. 7 we summarize our findings and  Appendices A-D provide further details in the derivation of certain results given in the text.

\section{Relativistic Particle-Field Dynamics}

The open quantum system paradigm starts by considering a system,
or universe, which is partitioned (according to natural physical
arguments or some large scale separation) into two smaller
systems. One subsystem, called the {\it system}, is assumed to be
the one of interest and the other, called the {\it environment},
contains many more degrees of freedom. An accurate description of
the behavior of the system variables requires knowing the
influence from the environment due to their mutual interactions.
A less than accurate description of the overall influence of the
environment  can be obtained by introducing some coarse-graining
over the environmental variables.

%Measurements are typically done by observing various expectation values and moments of the system variables. Feynman and Vernon (FV) studied several quantum mechanical models for computing true expectation values (not transition amplitudes). In quantumf field theory, Schwinger and Keldysh (SK) introduced the closed-time-path (CTP) generating functional for computing expectation values of quantum fields. Hu and Calzetta (???) later merged these formalisms together for studying the real-time dynamics of expectation values of systems/environments involving quantum fields.

\subsection{Worldline Influence Functional}

In a coordinate system, assume at some initial time $t_i$ the quantum statistical state
of the combined system S (particle in position $z_i$) and
environment E (quantum field $ \varphi_i$) is described by a
density matrix $\rho (z_i, \varphi_i; z_i ^\prime, \varphi_i
^\prime; t_i )$. In practice, specifying such a state is non-trivial since one requires a time-like Killing vector to define positive frequency modes and hence a Hilbert space of states. If the spacetime admits an asymptotically flat region or is conformally flat then the initial state can be constructed. But a general spacetime may not admit a time-like Killing vector. Regardless, we will sidestep these issues by working at a formal level. At some $t_f > t_i$ the density matrix is evolved to
\begin{eqnarray} 
	&& \!\!\!\! \rho (z_f, \phi_f; z_f ^\prime, \phi_f ^\prime; t_f ) \non \\
	&& ~ = \int dz_i d\phi_i \int dz_i ^\prime d\phi_i ^\prime  K(z_f, \phi_f, t_f; z_i, \phi_i, t_i ) \non \\
	&& {\hskip0.25in} \times \rho (z_i, \phi_i; z_i ^\prime, \phi_i
^\prime; t_i ) \, K ^* ( z_f ^\prime, \phi_f ^\prime, t_f; z_i
^\prime, \phi_i ^\prime, t_i ) \non \\
\end{eqnarray}
where $K$ is the amplitude of the time-evolution operator
$\hat{U}(t_f, t_i) = \exp \{- \frac{i}{\hbar} \int _{t_i} ^{t_f}
dt \, \hat{H}_{S+E}[z, \phi] \}$ for the system plus environment
and has a path integral representation given by
\begin{eqnarray}
K(z_f, \phi_f, t_f; z_i, \phi_i, t_i )  = \int _{z_i, \phi_i}
^{z_f, \phi_f} {\cal D} z {\cal D} \phi \, e^{\frac{i}{\hbar}
S_{S+E} [z, \phi] }
\end{eqnarray}
The action describing the system plus environment can be written
as the actions for the system $S_S[z]$ and environment
$S_E[\phi]$ along with an interaction action $ S_{int} [z, \phi]
$ between them
\begin{eqnarray}
S_{S+E} [ z, \phi ] &=& S_S[z] + S_E[\phi] + S_{int} [z, \phi]
\end{eqnarray}
with
\begin{eqnarray}
    S_S[z] &=& - m_0 \int d\tau \, \sqrt{ - g_{\mu \nu} u^\mu u^\nu } \non \\
    S_E [\phi] &=& \frac{1}{2} \int d^4 x \, \sqrt{-g} \left( g^{\mu\nu} \partial _\mu \phi \partial _\nu \phi - \xi_R R \phi^2 \right) \non \\
    S_{int} [z, \phi] &=& \int d^4x \, j(x; z] \, \phi(x) \non \\
    &=&  - \frac{e}{4 \pi} \int d\tau \, \sqrt{ - u^\mu u_\mu } \, \phi(z(\tau))
\end{eqnarray}
where the current density $j(x; z]$ is given by
\begin{eqnarray}
j(x; z] = - \frac{e}{4 \pi} \int d\tau \, \sqrt{- g_{\mu \nu}
u^\mu u^\nu} \, \frac{ \delta^4 (x - z(\tau)) }{ \sqrt{-g} }
\end{eqnarray}
At this stage, $\tau$ is just a parameter of the worldline and not necessarily the proper time.
We use an overdot to denote differentiation with respect to the
worldline parameter $\tau$. For example, the 4-velocity of a
particle in Minkowski space ($g_{\mu\nu}=\eta_{\mu\nu}$) is
$u^\alpha = \frac{d z^\alpha}{d\tau} = \dot{z}^\alpha$.
Because the particle is moving in a gravitational field, one
should replace ordinary derivatives $d/d\tau$ by covariant
derivatives $D/d\tau$ and so, for instance, the 4-acceleration is
$\frac{D u^\alpha}{d\tau} = u^\beta \nabla _\beta u^\alpha$ and
not $\dot{u}^\alpha = \frac{d u^\alpha}{d\tau} = u^\beta
\partial_\beta u^\alpha$.

To facilitate easier computation it is customary to choose the
initial density matrix to correspond to a factorized state of the
system and environment. Physically, this means that all of the field modes have been uncorrelated with the particle by an instantaneous measurement at time $t_i$. Aside from issues about performing this measurement simultaneously in the spacetime, this choice is somewhat unphysical
because, as explained in \cite{HPZ}, an infinite amount of
energy is required to uncorrelate all the modes of the
environment (field) from the system (particle) at a particular
instant of time. For instance, in models of quantum
Brownian motion with an infinite number of environment oscillators the factorized initial state results in large
transients of the diffusion coefficients appearing in the master
equation for the reduced density matrix \cite{HPZ}. The transients appear as a result of the high frequency modes of the environment beginning to interact and correlate with the system just after the initial time. This recorrelation time lasts on the order of the inverse of the cut-off frequency used to regulate the divergences coming from oscillators of very high
frequencies. For times much longer than this transient time the behavior due to the initial
factorization is usually discounted.  Other methods, including
the preparation function method \cite{Grabert_Schramm_Ingold}, allow for a somewhat
more physical initial state by including certain system-environment correlations (e.g. system in thermal equilibrium with the environment at the initial time), but still seem to suffer from some of the problems associated with the factorized state \cite{Romero_Paz}.

%Whatever coordinates we have chosen to label the spacetime
Having said this we assume that there is a Cauchy hypersurface at the initial time
$t_i$ such that
%an instantaneous measurement everywhere on the hypersurface uncorrelates the field
%modes from the particle variables so that
the initial density matrix takes the factorized form
\begin{eqnarray}
\rho (z_i, \phi_i; z_i ^\prime, \phi_i ^\prime; t_i ) = \rho_S
(z_i, z_i ^\prime; t_i ) \otimes \rho_E (\phi_i, \phi_i ^\prime;
t_i )
\end{eqnarray}
This simple form eases the manipulations for obtaining a description of the reduced particle dynamics.

After tracing out (a form of coarse-graining)  the field variables
from the density matrix the {\it reduced density matrix} for the system
is given by
\begin{widetext}
\begin{eqnarray}
    && \rho_r (z_f, z_f ^\prime; t_f ) = \int d\phi_f \, \rho  (z_f, \phi_f; z_f ^\prime, \phi_f; t_f ) \non \\
    && ~ = \int dz_i \, dz_i ^\prime \int _{z_i} ^{z_f} {\cal D}z \int _{z_i ^\prime} ^{z_f ^\prime} {\cal D} z^\prime \, \rho _S (z_i, z_i ^\prime; t_i ) \, e^{\frac{i}{\hbar} ( S_S [z] - S_S [z ^\prime] ) }  \non \\
    && {\hskip0.5in} \times \int d\phi_f \, d\phi_i \, d\phi_i ^\prime \int _{\phi_i} ^{\phi_f} {\cal D} \phi \int _{\phi_i ^\prime} ^{\phi_f } {\cal D} \phi ^\prime\rho_E (\phi_i, \phi_i ^\prime; t_i ) \, e^{\frac{i}{\hbar} ( S_E [\phi] + S_{int} [z, \phi] - S_E [\phi^\prime] - S_{int} [ z^\prime, \phi^\prime] ) } \non \\
    && ~ = \int dz_i \, dz_i ^\prime \int _{z_i} ^{z_f} {\cal D}z \int _{z_i ^\prime} ^{z_f ^\prime} {\cal D} z^\prime \,  \rho _S (z_i, z_i ^\prime; t_i ) \, e^{\frac{i}{\hbar} ( S_S [z] - S_S [z ^\prime] ) } \, F[z, z^\prime]  \label{red_rho}
\end{eqnarray}
The last line introduces the influence functional $F[z,
z^\prime]$, which is given by
\begin{eqnarray}
F[z, z^\prime] &=& \int d\phi_f \, d\phi_i \, d\phi_i ^\prime \int _{\phi_i} ^{\phi_f} {\cal D} \phi \int _{\phi_i ^\prime} ^{\phi_f } {\cal D} \phi ^\prime \,  \rho_E (\phi_i, \phi_i ^\prime; t_i ) \, e^{\frac{i}{\hbar} ( S_E [\phi] + S_{int} [z, \phi] - S_E [\phi^\prime] - S_{int} [ z^\prime, \phi^\prime] ) } = e^{\frac{i}{\hbar} S_{inf}[z, z^\prime] } \non \\
\end{eqnarray}
\end{widetext}
and $S_{inf}$ is the influence action. In operator language $F$ is
\begin{eqnarray}
F[z, z^\prime]  =  {\rm Tr}_{E} \: \hat{U}_{E+int} (t_f, t_i; z]
\: \hat{\rho}_{E} (t_i) \: \hat{U}^\dagger _{E+int} (t_f, t_i;
z^\prime] \non \\
&& \label{IF_op}
\end{eqnarray}
where $\hat{U}_{E+int}(t_f, t_i; z]$ is the evolution operator that evolves the environment variables through its interaction with the system. The influence functional can be interpreted as the overlap of the environment states evolved forward and backward in time while interacting with different particle trajectories, or simply as the ensemble average of the evolution operator $U_{E+int}[z]$ evolved backward in time under a different external source $z^\prime$, as can be seen by writing (\ref{IF_op}) as
\begin{eqnarray}
    && F[z, z^\prime] = \langle \hat{U} ^\dagger _{E+int} [z^\prime] \, \hat{U}_{E+int} [z] \rangle_{ens} \non \\
        && ~= \sum_\alpha \sum_{\alpha ^\prime} \rho _{E, \alpha \alpha^\prime} (t_i) \, \langle \alpha^\prime |
        \hat{U}_{E+int} ^\dagger [z^\prime] \, \hat{U} _{E+int} [z] | \alpha \rangle \non \\
\end{eqnarray}
In the interaction picture, the time evolution operator is
$\hat{U}_{E+int} (t_f, t_i; z] =  T e^{\frac{i}{\hbar}
\hat{\phi}_I \cdot j[z]}$ where the $\cdot$ denotes spacetime
integration so that for two functions, possibly tensors,
$A(x^\alpha)$ and $B(x^\beta)$
\begin{eqnarray}
    A \cdot B \equiv \int _{x^0 _i} ^{x^0 _f} dx^0 \int d^3 x \: \sqrt{-g} \: A(x) \: B(x)
\end{eqnarray}
Assuming that the initial state of the field is Gaussian, the influence functional can be calculated exactly giving
\begin{eqnarray}
    F[ z, z^\prime] &=& e^{-\frac{1}{4 \hbar} j^- \cdot (16 \pi^2 G_H ) \cdot j^- + \frac{i}{\hbar} \: j^- \cdot (16 \pi^2 G_{ret} ) \cdot j^+ } \non \\
\label{IF}
\end{eqnarray}
for the influence functional and
\begin{eqnarray}
    && S_{IF} [z, z^\prime] \bydefn -i \: \hbar \: {\rm ln} \: F[z, z^\prime] \non \\
        && ~= \frac{i}{4} \, j^- \cdot ( 16 \pi^2 G_{H} ) \cdot j^- + j^- \cdot ( 16 \pi^2 G_{ret} ) \cdot j^+ \non \\
 \label{S_inf}
\end{eqnarray}
for the influence action. The difference and semi-sum current densities are defined as
\begin{eqnarray}
    j^- &=& j[z] - j[z^\prime] \\
    j^+ &=& \frac{j[z] + j[z^\prime]}{2}
\end{eqnarray}
and the Hadamard $G_H$ and retarded Green's functions $G_{ret}$ are
\begin{eqnarray}
    && G_H ( x, x^\prime ) = \big\langle \{ \hat{\phi}_I (x), \hat{\phi}_I (x^\prime) \} \big\rangle - 2 \: \big\langle \hat{\phi}_I (x) \big\rangle \big\langle \hat{\phi}_I (x^\prime) \big\rangle \non \\
    && G_{ret} (x, x^\prime ) = i \, \theta_+ ( x, \Sigma ) \: \big\langle [ \hat{\phi}_I (x), \hat{\phi}_I (x^\prime) ] \big\rangle   \non \\
       \label{ret}
\end{eqnarray}
The $\langle \cdots \rangle = {\rm Tr}_E \, \hat{\rho}_E
(\cdots)$ denotes quantum expectation values in the Gaussian
initial state $\hat{\rho}_E$ of the environment. The step
function $\theta_+ (x, \Sigma)$ appearing in $G_{ret}$ equals one
in the future of the point $x^\prime$ and zero otherwise. Here,
$\Sigma$ is a space-like hypersurface containing $x^\prime$. Had
the initial state contained non-Gaussian contributions, one would
have many additional terms involving cubic and higher powers of
the coupling. Likewise for nonlinear interactions (e.g. $\sim e
j[z] \cdot \phi^n$). Assuming that the coupling is small and that
non-Gaussianities are also small one can still use (\ref{IF}) and
(\ref{S_inf}) as a lowest order approximation.

\subsection{Semi-Classical Particle Dynamics}

The reduced density matrix for the particle is now written as
\begin{eqnarray}
    && \rho_r (z_f, z_f ^\prime; t_f ) \non \\
    && ~= \!\! \int dz_i \, dz_i ^\prime \int _{z_i} ^{z_f} \!\! {\cal D} z \int _{z_i^\prime} ^{z_f ^\prime} \!\! {\cal D} z^\prime \, \rho_S (z_i, z_i^\prime; t_i ) \, e^{ \frac{i}{\hbar} S_{CGEA} [z, z^\prime] } \non \\
\end{eqnarray}
where the {\it coarse-grained effective action} (CGEA) is defined as
\begin{eqnarray}
    && \!\!\!\! S_{CGEA} [z, z^\prime] \non \\
    &&~= S_S [z] - S_S [z^\prime] + S_{IF} [z, z^\prime] \non \\
        && ~= S_S [z]  - S_S [z^\prime] + j^- \cdot ( 16 \pi^2 G_{ret} ) \cdot j^+ \non \\
        && {\hskip0.25in} + \frac{i}{4} \, j^- \cdot ( 16 \pi^2 G_{H} ) \cdot j^-
\end{eqnarray}
At this point it is worth mentioning that the magnitude of the influence functional decays rapidly for two largely separated histories since it is Gaussian (to lowest order) in $z^- = z-z^\prime$
\begin{eqnarray}
    && | F[z,z^\prime] | = e^{-\frac{1}{4 \hbar} j^- \cdot (16 \pi^2 G_H ) \cdot j^-} \non \\
        && ~= \exp -\frac{e^2}{4 \hbar} \int d\tau \int d\tau^\prime \, z_- ^{\alpha} \, \frac{\delta j^-}{\delta z_-^{\alpha} } \, G_H (z_+^\mu, z_+^{\mu^\prime} ) \, \frac{\delta j^-}{\delta z_-^{\alpha^\prime} } \, z _-^{\alpha^\prime}  \non \\
        && {\hskip0.25in} + {\cal O} ( z_-^4 )
\end{eqnarray}
Here, and in the following, $z^\alpha \equiv z^\alpha (\tau)$ and $z^{\alpha^\prime} \equiv z^\alpha (\tau^\prime)$ so that an unprimed (primed) index refers to that component of a tensor field or coordinate evaluated at proper time $\tau$ ($\tau^\prime$) and
\begin{eqnarray}
    z_\mu ^- &=& z_\mu - z^\prime _\mu \\
    z_\mu^+ &=& \frac{ z_\mu + z^\prime _\mu }{2}
\end{eqnarray}
The norm of $F$ is equal to the norm of the decoherence
functional (see below), which is a measure of how much the
particle's worldline is decohered. So, if the quantum
fluctuations of the field (environment) provide a strong enough
mechanism for decoherence (this should be checked on a case by
case basis and will be assumed true in the cases under study
here) then we are justified in expanding the CGEA about the
classical trajectory $\bar{z}^\mu$. Doing so gives
\begin{eqnarray}
    S_{CGEA} [z,z^\prime] = \int d\tau^\prime \, z_- ^{\alpha^\prime} \, \frac{\delta S_{CGEA} }{ \delta z_- ^{\alpha^\prime} } \bigg| _{z = z^\prime = \bar{z} } + {\cal O} (z_- ^2 ) \non \\
\end{eqnarray}
Using this expression in the
reduced density matrix and doing a stationary phase approximation
gives the equations of motion for the classical worldline
$\bar{z}$
\begin{eqnarray}
    \frac{ \delta S_{CGEA} }{ \delta z_- ^\mu (\tau) } \bigg| _{z=z^\prime = \bar{z} } = 0
\label{CGEA_funcvar}
\end{eqnarray}
Evaluating the functional derivative using, for some test function $f(x)$,
\begin{eqnarray}
    && \frac{\delta}{\delta z_- ^\mu (\tau) } \int d^4x \, \sqrt{-g} \, j^- (x; z] f(x) \non \\
    && {\hskip0.25in} = \frac{e}{4\pi} \, \left( \frac{D u_\mu}{d\tau} + w_\mu ^{~ \nu} [z] \nabla_{z^\nu} \right) f(z) \non \\
    && {\hskip0.25in} \bydefn \frac{e}{4\pi} \, \vec{w}_\mu [z] \, f(z)
\end{eqnarray}
where $w_\mu ^{~ \nu} = g_\mu ^{~\nu} + u_\mu u^\nu$ and $u^\alpha u_\alpha = -1$ (proper-time gauge), gives
\begin{eqnarray}
    m_0 \, \frac{D \bar{u}_\mu}{d\tau} = e \, \vec{w}_\mu [\bar{z}] \phi_{ret} (\bar{z})
\label{mean_eom}
\end{eqnarray}
where $\phi_{ret}$ is the retarded field
\begin{eqnarray}
    \phi_{ret} (x) = e \int d\tau^\prime \, G_{ret} (x, z^{\alpha^\prime})
\label{ret_field}
\end{eqnarray}
The right side of (\ref{mean_eom}) is the self-force on the
particle arising from the radiation reaction. Furthermore, the
vector operator $\vec{w}_\mu$ contains the particle's
acceleration implying that the particle moves with an effective
time-dependent mass equal to $m_0 - e \, \phi_{ret} (\bar{z})$.
(See \cite{Burko_Harte_Poisson} for a discussion of evaporating
scalar charges based on this time-dependent effective mass.)
Provided there is strong enough decoherence to suppress the
quantum fluctuations of the worldline, the equations of motion
for the quantum expectation value $\langle \hat{z}^\mu \rangle$
are the same as in (\ref{mean_eom}) at tree-level in both the
particle and the field \cite{JH1}. Unfortunatley,
(\ref{mean_eom}) is problematic since the self-force diverges on
the worldline of the point particle. This divergence can be
traced to the singular coincidence limit of the retarded Green's
function $\lim_{\tau^\prime \rightarrow \tau} G_{ret} \rightarrow
\infty$ and results from the point particle assumption. We will
discuss how to treat this problem in Sec. \ref{Reg} by adopting
an effective field theory point of view.

\subsection{Stochastic Semi-Classical Particle Dynamics}

We have made the assumption that the quantum fluctuations of the
worldline are strongly suppressed by the decoherence due to the
quantum field. Even under strong decoherence when the classical
trajectory is well-defined, the quantum fluctuations of the field
can still influence the classical motion of the particle
%by exchanging momentum and energy
through the particle-field coupling $S_{int}$. They may show up as
classical stochastic forces on the particle. In this section we
shall show how this comes about using the influence functional
for the reduced density matrix.

We start by invoking the relation
\begin{eqnarray}
    && e^{-\frac{1}{4 \hbar} j^- \cdot ( 16 \pi^2 G_H) \cdot j^-}  = N \!\! \int {\cal D} \xi(x) \, e^{- \frac{1}{\hbar} \xi \cdot G_H ^{\, -1} \cdot \xi - \frac{ 4 \pi i }{ \hbar} \xi \cdot j^-} \non \\
\end{eqnarray}
where $N$ is a normalization factor that is independent of the worldline coordinates and $\xi(x)$ is some auxiliary field. Using this relation, the influence functional (\ref{IF}) can be written as
\begin{eqnarray}
    F[z, z^\prime] = N \!\! \int \!\! {\cal D} \xi(x) \, e^{- \frac{1}{\hbar} \xi \cdot G_H ^{\, -1} \cdot \xi + \frac{ i }{ \hbar}  j^- \!\! \cdot ( - 4 \pi \xi + 16 \pi^2 G_{ret} \cdot j^+ ) } \non \\ 
\end{eqnarray}
where $\xi(x)$ appears as an auxiliary function, which can be
interpreted as a classical stochastic or noise field \cite{JH1, Calzetta_Roura_Verdaguer}
with an associated Gaussian probability distribution functional
\begin{eqnarray}
P_\xi [\xi(x)] = e^{-\frac{1}{\hbar} \xi \cdot G_H ^{\, -1} \cdot \xi }
\end{eqnarray}
The fact that this is Gaussian is a direct consequence of taking an initial Gaussian state for the quantum field. With respect to $P_\xi [\xi]$ this implies that $\xi$ has zero-mean
and its correlator is proportional to the Hadamard function
encoding the information about the fluctuations in the quantum
field \cite{footnote_1}. %\footnote{In general, for systems with a nonlinear coupling to the environment or for non-Gaussian initial states of the environment the interpretation of $P_\xi$ as a probability distribution is not always possible. It may take on negative values in which case $P_\xi$ should be interpreted as a pseudo-probability distribution in a similar vein as the Wigner function \cite{Calzetta_Roura_Verdaguer}.}
\begin{eqnarray}
    \big\langle \xi(x) \big\rangle_\xi &=& 0 \\
    \big\langle \{ \xi(x), \xi(x^\prime) \} \big\rangle _\xi &=& \hbar \, G_H (x, x^\prime)   \label{xi_correlator}
\end{eqnarray}
where $\langle \ldots \rangle _\xi = N \int {\cal D} \xi \, P_\xi \, (\ldots)$.

Now the reduced density matrix (\ref{red_rho}) becomes
\begin{widetext}
\begin{eqnarray}
    \rho_r (z_f, z_f ^\prime; t_f) = N \int dz_i \, dz_i ^\prime \int _{z_i} ^{z_f} \!\! {\cal D} z \int_{z_i ^\prime} ^{z_f^\prime} \!\! {\cal D} z^\prime \, \rho_S (z_i, z_i ^\prime; t_i ) \int {\cal D} \xi \, P_\xi [\xi] \, e^{ \frac{i}{\hbar} S_{SEA} [z, z^\prime; \xi] }
    \label{rho_sea}
\end{eqnarray}
where the {\it stochastic effective action} (SEA) is defined as
\begin{eqnarray}
    S_{SEA} [z, z^\prime; \xi] = S_S [z] - S_S [z^\prime] + j^- \! \cdot ( - 4 \pi \xi + 16 \pi^2 G_{ret} \cdot j^+ ) = S_{CGEA} ^R [z,z^\prime] - (4 \pi) \, \xi \cdot j^-
\end{eqnarray}
and $S_{CGEA}^R$ is the real part of the CGEA. As before, if the worldline is strongly decohered by the quantum fluctuations of the environment then we are justified in expanding the SEA around the classical solution
\begin{eqnarray}
    S_{SEA} [z, z^\prime; \xi] = - \int d\tau^\prime z_- ^{\alpha^\prime} \eta_{\alpha^\prime} [z^+] + \frac{1}{2} \int d\tau^\prime \int d\tau^{\prime \prime} \, z_- ^{\alpha^\prime} z_- ^{\alpha^{\prime\prime}} \frac{ \delta^2 S_{CGEA}^R }{ \delta z_+ ^{\alpha^\prime} \delta z_+ ^{\alpha ^{\prime \prime} } } \bigg| _{z=z^\prime = \bar{z}} + {\cal O} (z_- ^3)
\label{SEA_expansion}
\end{eqnarray}
\end{widetext}
where we have used (\ref{CGEA_funcvar}), expanded $j^-$ in powers of $z_-$ and defined the stochastic force
\begin{eqnarray}
    \eta_\mu [z] = e \, \vec{w}_\mu[z] \xi(z)
\end{eqnarray}
Putting (\ref{SEA_expansion}) into the reduced density matrix and using the stationary phase approximation gives the stochastic equations of motion for the worldline fluctuations $z_- ^\mu = \tilde{z}^\mu \equiv z^\mu - \bar{z}^\mu$
\begin{eqnarray}
    \int d\tau^\prime \, \tilde{z}^{\alpha^\prime} \frac{ \delta^2 S_{CGEA}^R }{ \delta \bar{z} ^\mu \delta \bar{z} ^{\alpha ^\prime } } \bigg| _{z=z^\prime = \bar{z}} = \eta_\mu [\bar{z}]
\label{z_flucs_funcder}
\end{eqnarray}
It should be emphasized that this equation describes the
evolution of small perturbations $\tilde{z}$ around the
semi-classical trajectory that arise from the stochastic
manifestation $\eta$ of the quantum field fluctuations.
%Care needs to be taken in evaluating the functional derivatives since it contains divergences. We will postpone doing this until section 5.
We can obtain a stochastic version of the ALD equation (\ref{mean_eom}) if we add the equations of motion for the classical dynamics
\begin{eqnarray}
    0 = \frac{\delta S_{CGEA}^R }{ \delta z_- ^\mu } \bigg|_{z=z^\prime=\bar{z}}
\end{eqnarray}
to the left side. Then we may write
\begin{eqnarray}
    \frac{\delta S_{CGEA}^R }{ \delta z_- ^\mu } \bigg|_{z_-=0} = \eta_\mu [z]
\end{eqnarray}
Evaluating this functional derivative gives the stochastic semi-classical particle dynamics for the full worldline $z_\mu = \bar{z}_\mu+\tilde{z}_\mu$
\begin{eqnarray}
    m_0 \, \frac{Du_\mu}{d\tau} = e \, \vec{w}_\mu [z] \phi_{ret} (z) + \eta_\mu [z]
\label{stoch_eom}
\end{eqnarray}
Notice that both the deterministic and the stochastic components
of the self-force can push the particle away from its geodesic
motion with respect to a fixed background spacetime. However, we should
keep in mind that this equation is really only valid to linear
order in the fluctuations $\tilde{z}$ since higher order terms
will correspond to quantum corrections that we have been
neglecting. Of course, we could have obtained (\ref{stoch_eom})
by using a stationary phase approximation in the reduced density
matrix (\ref{rho_sea}) without having expanded around the classical trajectory.
However, we believe that the validity of (\ref{stoch_eom}) to
linear order in the fluctuations would not have been as
transparent.

The stochastic correlation functions of the force $\eta_\mu$ can
be evaluated using the knowledge of the $\xi$ correlators above.
As was commented in the previous paragraph we must evaluate these
correlation functions along the classical trajectory $\bar{z}$ to be consistent with the linearization.
The mean of the stochastic force is zero
\begin{eqnarray}
    \big\langle \eta_\mu [\bar{z}] \big\rangle _\xi = e \, \vec{w}_\mu [\bar{z}] \, \big\langle \xi (\bar{z}) \big\rangle _\xi = 0
\end{eqnarray}
and the symmetric two-point function of the stochastic force is
\begin{eqnarray}
    \big\langle \{ \eta_\mu [ \bar{z} ], \eta_{\mu^\prime} [ \bar{z} ] \} \big\rangle _\xi =  \hbar \, e^2 \, \vec{w}_{\left( \mu \right. } [\bar{z}] \, \vec{w}_{\left. \mu^\prime \right) } [\bar{z}] \, G_H (\bar{z}^\alpha, \bar{z}^{\alpha^\prime} ) \non \\
\label{eta_corr}
\end{eqnarray}
This shows that the noise $\eta_\mu$ is multiplicative, colored
and depends on the particle's initial conditions through the
classical trajectory. The noise correlator also generically
depends on the field's initial conditions as is seen by the
appearance of $G_H$ in the equation. The Hadamard function $G_H$
does not vanish on a space-like hypersurface implying that the
quantum correlations in the environment are non-local. However,
the noise correlator above is evaluated on time-like separated
points only. For equal proper times $\tau^\prime = \tau$ the
Hadamard function diverges so a suitable regularization procedure
must be used in order to make sense of (\ref{eta_corr}) near
coincidence.

%As mentioned earlier, the retarded Green's function $G_{ret}$ also contains a divergence resulting from the self-force evaluated along the point particle worldline.

%Notice that the noise $\eta_\mu (\tau; z]$ depends on both the
%proper time and the worldline coordinates of the particle, which
%is to be determined, making this a very difficult equation to
%solve. However, since the fluctuations are of order $\hbar$ we
%expect the particle not to be perturbed far away from its mean
%trajectory. So instead of attempting to find exact solutions of
%(\ref{stoch_eom}), we will linearize this equation around the
%mean trajectory (using (\ref{mean_eom})) to get equations for the
%worldline fluctuations (along a prescribed mean trajectory so
%that $\eta_\mu$ behaves as an external force independent of the
%fluctuations.) Given a solution to (\ref{mean_eom}) one may then
%solve for the fluctuations (numerically likely) and determine if
%and when the linear approximation breaks down.

In the next two sections we describe a new method introduced in
\cite{JH1} for regulating the ultraviolet divergences in the
retarded Green's function. We introduce a high energy scale in
the quantum field below which the low energy point particle dynamics is
expected to be valid. For long times as compared to the inverse
of the high energy scale, we invoke a quasi-local expansion in
order to obtain the relevant contributions to the particle's
motion. Our work is a generalization of \cite{JH1} to curved
spacetime. We follow Poisson's description of scalar radiation
reaction in curved spacetime and use units where $c = G = 1$. (We
change the particle physics notation in \cite{JH1} (+, -, -, -)
to the MTW notation (-, +, +, +) of \cite{Poisson}.)

\section{Regularization of the Retarded Green's Function}
\label{Reg}

Before we discuss regularizing the retarded Green's function it will be beneficial to introduce some notation. Following \cite{Poisson} we define the step function $\theta_+(x, \Sigma)$ to be 1 for any point $x$ to the future of a space-like hypersurface $\Sigma$ and $0$ otherwise. This is a generalization of the ordinary step function to curved spacetime. Using this, one can define the following distributions
\begin{eqnarray}
    \delta _+ ( \sigma(x,x ^\prime) ) &=& \theta_+ (x, \Sigma) \: \delta ( \sigma (x, x^\prime) ) \\
    \theta_+ ( -\sigma( x, x^\prime ) ) &=& \theta_+ ( x, \Sigma) \: \theta ( - \sigma ( x, x^\prime ) )
\end{eqnarray}
where $\sigma (x, x^\prime)$ is Synge's world function along the (unique) geodesic linking  $x=z(\tau)$ and $x^\prime=z(\tau^\prime)$. Taking the lightcone centered on $x^\prime$, with $x^\prime$ on the space-like hypersurface $\Sigma$, it follows that $\delta_+(\sigma)$ has support along the forward lightcone only while $\theta_+ (-\sigma)$ is one in the causal future of $x^\prime$ and vanishes everywhere else.

In (\ref{ret}) the retarded Green's function is calculated from
the commutator of the interaction picture field $\hat{\phi}_I$.
Recall that $\hat{\phi_I}$ evolves under the free dynamics so that
\begin{eqnarray}
    (\Box - \xi_R R) \, \phi_I = 0
\end{eqnarray}
and so is the homogeneous solution to the full field equation in
the Heisenberg picture (satisfying the same initial conditions)
\begin{eqnarray}
    (\Box - \xi_R R) \, \phi_H = j[z]
\end{eqnarray}
where $\hat{\phi}_H = \hat{\phi}_I + \hat{1} \, G_{ret} \cdot
j[z]$ for a classical source. Because the source is classical
only the first term in the solution gives a contribution to the
commutator and hence to the retarded Green's function
\begin{eqnarray}
    G_{ret} \sim \big\langle [ \hat{\phi}_H (x), \hat{\phi}_H (x^\prime) ] \big\rangle = \big\langle [ \hat{\phi}_I (x), \hat{\phi}_I (x^\prime) ] \big\rangle
\end{eqnarray}
Finally, since the commutator is state-independent it follows
that $G_{ret}$ for the quantum field $\hat{\phi}_H$ and for the
corresponding classical field $\phi$ are the same.

\subsection{Hadamard Expansion}

This allows us to invoke Hadamard's ansatz for which the retarded
Green's function for a scalar field in a curved $3+1$ dimensional
spacetime has the form \cite{Poisson}
\begin{eqnarray}
    && G_{ret} ( x, x^\prime ) \non \\
    && ~= \Delta ^{1/2} ( x, x^\prime ) \, \delta_+ ( \sigma (x, x^\prime ) ) + V( x, x^\prime  ) \: \theta_+ ( -\sigma (x, x^\prime  ) ) \non \\
\label{Hadamard}
\end{eqnarray}
which is the sum of a ``direct" part (proportional to $\delta_+$)
and a ``tail" part (proportional to $\theta_+ (-\sigma)$). Notice
that the appearance of $\delta_+$ and $\theta_+$ ensures that
$G_{ret}$ has the correct causal structure. In order for this
ansatz to be valid, the points $x$ and $x^\prime$ must be
connected by a unique geodesic. Otherwise, ambiguities arise from
the appearance of caustics when parallel propagating a tensor
field from $x^\prime$ to $x$ or vice versa. This
will be assumed in the following. The function $\Delta (x,x
^\prime)$ is the van Vleck determinant and the function $V(x,
x^\prime)$ satisfies the homogeneous Klein-Gordon equation in
curved spacetime
\begin{eqnarray}
    ( \Box_g - \xi_R R ) \, V(x, x^\prime) = 0
\end{eqnarray}
with boundary data determined by the restriction of $V$ to the forward lightcone, denoted by $\tilde{V}$. The restriction of $V$ is found by solving
\begin{eqnarray}
    \tilde{V}_{,\alpha} \sigma^\alpha + \frac{1}{2} \left( \sigma ^\alpha _{~\alpha} -2 \right) \tilde{V} = \frac{1}{2} \left\{ \left( \Box_g - \xi_R R \right) \Delta^{1/2} \right\}_{\sigma=0} \non \\
\label{V_tilde}
\end{eqnarray}
Knowing all of the light-like geodesics emanating into the future
from $x^\prime$, one can integrate this equation to construct
$\tilde{V}$ on the light-cone and hence $V$ inside the
light-cone. Of course, this just solves for $V(x,x^\prime)$ for
any $x$ to the causal future of $x^\prime$ and must be solved
again for different values of $x^\prime$. This makes determining $V$ very
difficult and tedious for generic spacetimes. However, if the spacetime
possesses enough symmetry (e.g. de Sitter, some FRW cosmologies) then $V$ can be
constructed relatively easily \cite{Burko_Harte_Poisson}.

From the mean and stochastic equations of motion (\ref{mean_eom}) and (\ref{stoch_eom}) it is necessary to compute the restriction of $G_{ret}$ to the particle's worldline
\begin{eqnarray}
    && G_{ret} ( z^\alpha, z ^{\alpha ^\prime} ) = \Delta ^{1/2} ( z^\alpha, z^{\alpha^\prime} ) \, \delta_+ ( \sigma (z^\alpha, z^{\alpha^\prime} ) ) \non \\
    && {\hskip 0.5in} + V( z^\alpha, z^{\alpha^\prime}  ) \: \theta_+ ( -\sigma (z^\alpha, z^{\alpha^\prime}  ) ) 
\label{Hadamard_Ansatz}
\end{eqnarray}
Using the Hadamard ansatz and (\ref{ret_field}) it is easy to see
that for two (time-like separated) points on the particle
worldline, $\delta_+(\sigma)$ has support only at coincidence,
when $\tau^\prime = \tau$. The contribution from the direct term
is then
\begin{eqnarray}
    && \int _{\tau_i} ^{\tau_f} d\tau^\prime \, \Delta^{1/2} (z^\alpha, z^{\alpha ^\prime} ) \, \theta_+( z^\alpha, S) \, \delta ( \sigma ( z^\alpha, z^{\alpha^\prime} ) ) \non \\
    && \!\!\!\! = \int _{\tau_i} ^\tau d\tau^\prime \, \Delta^{1/2} (z^\alpha, z^{\alpha^\prime} ) \, \delta ( \tau^\prime - \tau ) \, \left( \frac{ d\sigma }{ d\tau } \right) _{\tau^\prime = \tau} ^{-1} 
    \label{divergence}
\end{eqnarray}
This expression is ultraviolet divergent since $d\sigma/d\tau$ evaluates to zero for $\tau^\prime = \tau$ and so the retarded field is not
well-defined (in fact, it is infinite) when evaluated at the
location of the point particle. This feature is not limited to just a
quantum field as it appears even for a classical field since they
share the same retarded Green's function.

The presence of this divergence suggests that a regularization
procedure is needed to render the equations of motion
(\ref{mean_eom}) and (\ref{stoch_eom}) finite. Several procedures
have been proposed in the literature for doing this. In
\cite{Detweiler_Whiting, Poisson}, the retarded field is evaluated
near the worldline so that the divergence can be tracked in the
limit that one approaches the worldline. This divergence
renormalizes the particle's mass. Writing the retarded field in
terms of functions singular $\phi_S$ and regular $\phi_R$ on the
worldline allows, after calculating their gradient and expanding
them near the worldline, for the self-force to be evaluated
unambiguously along the worldline. This method can also be used
for electromagnetic and gravitational self-force calculations. In
another method developed in \cite{Barack_Ori, Barack, BMNOS,
Barack_Ori_2, Mino_Nakano_Sasaki}, a mode-sum regularization
procedure is used in which the self-force is decomposed in terms
of the multipole moments of the field. From this one subtracts
the moments of the singular part of the self-force and resums
over the multipoles to obtain a self-force that is finite on the
worldline. In \cite{Rosenthal} the ``massive field approach" is
used in which an auxiliary massive field is introduced to
subtract away the ultraviolet divergence from the massless field.
This procedure is similar to Pauli-Villars type regularization
and has also been used in \cite{JohnsonPhD} to regularize the
direct term in $G_{ret}$.

Other regularization methods have been used for the radiation
reaction of non-relativistic particles without gravity present
\cite{NR_Reg}. Notable of these include the extended-charge
models of \cite{Ext_Charge} in which the particle is described as
a rigid volume with a non-zero size. Unfortunately, this type of
prescription cannot be extended to relativistic models since
rigid bodies are incompatible with relativity \cite{footnote_2}. %\footnote{Recently, the authors of \cite{Goldberger_Rothstein} have used the methods of effective field theory to describe the gravitational two-body problem of extended objects as an effective point-particle theory.}. 
Furthermore, the successes of quantum field theory
(e.g. as applied to the scattering of point particles), has
proved to be an adequate description of the low-energy effective
dynamics of point particles to $\sim$TeV scales. It seems
unnecessary, then, to introduce non-trivial structure to the
particle to study the self-force.

\subsection{Regularization motivated by Effective Field Theory}

With this consideration  in mind we introduce yet another
regularization for the self-force motivated by \cite{JH1}. All
theories for the fundamental interactions (e.g. QED, electroweak
theory, QCD) can be interpreted as effective theories, in that
they are are low energy limits of a more complete theory valid at
higher energies, yet they provide an excellent description of low
energy phenomena,  at least for renormalizable theories, without
requiring the details of the complete theory. For energies much
higher than the energy scale of the effective theory, new physics
is likely to become important.

In this spirit we introduce a
regulator $\Lambda$ for the field. Above the energy scale of this
regulator, $E_\Lambda$, new physics is assumed to occur. For
energies much lower than $E_\Lambda$, the dynamics
(\ref{mean_eom}) and (\ref{stoch_eom}) will be sufficiently
accurately described by using a regulated quantum field. This
approach has been taken in \cite{JH1} in deriving
the Abraham-Lorenz-Dirac (ALD) equations in flat spacetime. It is easily extended to
motions in a curved spacetime since the direct part of the
retarded Green's function is a local quantity on the worldline.

Regularization is achieved here by choosing any suitably smooth
function to approximate $\delta_+$ for large values of the
regulator $\Lambda$. Following \cite{JH1} we replace $\delta_+$ with
\begin{eqnarray}
    \delta_+ (\sigma (z^\alpha, z^{\alpha^\prime}) ) &\rightarrow& \theta_+ (-\sigma(z^\alpha, z^{\alpha^\prime} ) ) \, \sqrt{ \frac{8}{\pi} } \, \Lambda^2 \, e^{-2 \Lambda^4 \sigma^2 (z, z^\prime) } \non \\
    &=& \theta_+ ( -\sigma ( z^\alpha, z^{\alpha^\prime} ) ) \, g_{ret} ^\Lambda ( z^\alpha, z^{\alpha^\prime} )
\label{replacement}
\end{eqnarray}
In the limit of infinitely large $\Lambda$ we have that $\theta_+ \, g_{ret}^\Lambda$ approaches $\delta_+$. The integral of the direct part of $G_{ret}$ (\ref{divergence}) receives a contribution from $\delta_+$ at coincidence implying that $\sigma = 0$. The function $g_{ret}^\Lambda$ is smooth but approximates $\delta_+$ well only if $\Lambda^2 \sigma \gg 1$. This approximation will not hold if $\sigma$ is identically zero. Nevertheless, we will assume that $\sigma$ is small and approaching zero but $\Lambda$ is such that $\Lambda^2 \sigma \gg 1$ still holds. This separation of scales allows us to do a quasi-local expansion below in which the self-force will be expanded near coincidence.

Making the replacement (\ref{replacement}) in $G_{ret}$ gives the regulated Green's function
\begin{widetext}
\begin{eqnarray}
    G_{ret} ^\Lambda (z^\alpha, z^{\alpha^\prime} ) &=& \theta_+ ( - \sigma( z^\alpha, z^{\alpha^\prime} ) ) \left\{ \Delta^{1/2} (z^\alpha, z^{\alpha^\prime} ) \, g_{ret} ^\Lambda ( z^\alpha, z^{\alpha^\prime} ) + V( z^\alpha, z^{\alpha^\prime} ) \right\} = \theta_+ ( -\sigma) \left\{ \Delta^{1/2} \, g_{ret}^\Lambda + V \right\}
\end{eqnarray}
where an abbreviated notation has been used on the far right side to ease the appearance of later expressions. Also, it should be clear that in the limit $\Lambda \rightarrow \infty$ that $G_{ret}^\Lambda \rightarrow G_{ret}$.

Putting $G_{ret}^\Lambda$ into (\ref{mean_eom}) gives
\begin{eqnarray}
    m_0 \, \frac{D \bar{u}_\mu}{d\tau} = e^2 \, \left( \frac{D \bar{u}_\mu}{d\tau} + w_\mu ^{~\nu} [\bar{z}] \, \nabla_{\bar{z}^\nu} \right) \int _{\tau_i} ^{\tau_f} d\tau^\prime \, \theta(\tau- \tau^\prime) \left\{ \Delta^{1/2} \, g_{ret}^\Lambda + V \right\}
\end{eqnarray}
where the quantities in the integral are evaluated along the classical trajectory $\bar{z}$. Passing the derivative through the integral gives
\begin{eqnarray}
    m_0 \, \frac{D \bar{u}_\mu}{d\tau} &=& e \, \frac{D \bar{u}_\mu}{d\tau} \, \phi_{ret} (\bar{z}) + e^2 \, w_\mu ^{~\nu} [\bar{z}] \left\{  \left[ \Delta^{1/2} g_{ret}^\Lambda + V \right] \left[ \nabla_{\bar{z}^\nu} ( \tau -\tau^\prime ) \right] + \int _{\tau_i} ^\tau d\tau^\prime \, \nabla_{\bar{z}^\nu} \left( \Delta^{1/2} g_{ret} ^\Lambda + V \right) \right\} \nonumber
\end{eqnarray}
where the $[ \ldots ]$ denotes the coincidence limit $\tau^\prime \rightarrow \tau$ of the quantity it contains. This notation looks the same when denoting a functional of a function (e.g. $w_\mu ^{~\nu} [z]$) but the context should be clear. From Appendix A one can show that
\begin{eqnarray}
    \left[ \nabla_{\bar{z}^\nu} ( \tau - \tau^\prime ) \right] = - \left[ \nabla_\nu (\bar{\sigma}^\alpha \bar{u}_\alpha) \right] =  - \bar{u}_\nu   \label{proof}
\end{eqnarray}
and since $w_\mu ^{~\nu} [\bar{z}]$ projects vectors onto a direction orthogonal to the mean 4-velocity then $w_\mu ^{~\nu} [\bar{z}] \left[ \nabla_{\bar{z}^\nu} ( \tau -\tau^\prime) \right]$ gives no contribution leaving
\begin{eqnarray}
    m_0 \, \frac{D \bar{u}_\mu}{d\tau} &=& e^2 \, \frac{D \bar{u}_\mu}{d\tau} \, \phi_{ret} (\bar{z}) + e^2 \, w_\mu ^{~\nu} [\bar{z}] \int _{\tau_i} ^\tau d\tau^\prime \, \nabla_{\bar{z}^\nu} \left( \Delta^{1/2} \, g_{ret} ^\Lambda + V \right) \\
    &=& e^2 \, \frac{D \bar{u}_\mu}{d\tau} \, \phi_{ret} (\bar{z}) + e^2 \, w_\mu ^{~\nu} [\bar{z}] \int _{\tau_i} ^\tau d\tau^\prime \, \left\{ \left(  \nabla_{\bar{z}^\nu} \, \Delta^{1/2} \right) g_{ret}^\Lambda + \Delta^{1/2} \,  \nabla_{\bar{z}^\nu} g_{ret}^\Lambda +  \nabla_{\bar{z}^\nu} V \right\}
\label{starting_pt}
\end{eqnarray}
\end{widetext}
It will be convenient for later manipulations to define the {\it tail term} as the integral over the past history of the gradient of $V$,
\begin{eqnarray}
    \phi_\nu ^{tail} (\bar{z}(\tau)) = e \int _{\tau_i} ^\tau d\tau^\prime \,  \nabla_{\bar{z}^\nu} V
\label{tail_term}
\end{eqnarray}
The tail term will turn out to be responsible for the non-Markovian and history-dependent dynamics of the particle's evolution. Consequently, this presents a significant source of difficulty when trying to determine the particle's motion since one has to know the entire past history of the particle's worldline in order to make predictions about its current position and speed.

\section{Effect of Radiation Reaction on Particle Trajectory}

In order to simplify some of the difficulties of solving the nonlinear integro-differential equation (\ref{starting_pt}) we utilize the effective theory viewpoint discussed in the previous section. Since $g_{ret}^\Lambda$ is strongly peaked around $\sigma = 0$ then the contributions to the self-force will be largest when $\tau^\prime \approx \tau$. We therefore introduce a quasi-local expansion by expanding $\theta_+ \, \Delta^{1/2} \, g_{ret}^\Lambda$ around coincidence $\tau^\prime = \tau$.
This yields a well-defined description of the self-force in terms of relevant and irrelevant quantities. The relevant terms are those that either renormalize certain parameters of the
particle or otherwise affect the particle dynamics when $\Lambda$
goes to infinity while the irrelevant terms are those that give no contribution in the same limit.

\subsection{Quasilocal Expansion}

\begin{figure}
	\includegraphics[height=6cm]{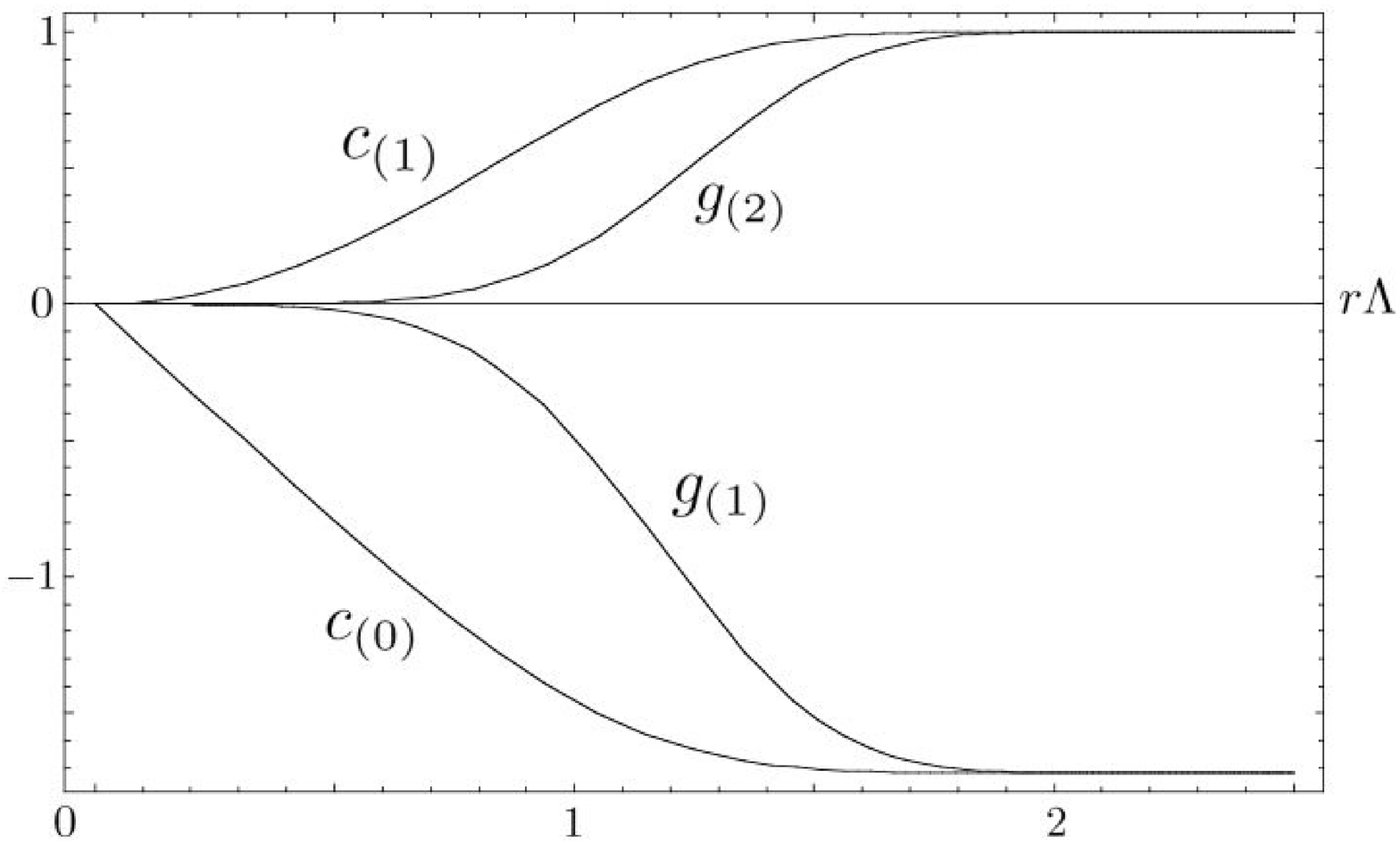}
    	\caption{Time-dependence of the coefficients appearing in (\ref{h})-(\ref{Aone}). The functions $c_{(0)}$ and $g_{(1)}$ have been divided through by $\Lambda$ so that they can be displayed along with $c_{(1)}$ and $g_{(2)}$.}
    \label{fig1}
\end{figure}

We begin by expanding the van Vleck determinant and
$g_{ret}^\Lambda$ for small values of $\sigma(z^\alpha, z^{\alpha^\prime})$. 
%It is important to recognize that this is not necessarily an expansion near coincidence since small values of $\sigma$ can be obtained for nearly light-like separated points even though they are separated by large time intervals and spatial distances. Nevertheless, we expand around small $\sigma$ and afterwards expand near coincidence.
%to guarantee that the local contribution of $G_{ret}$ is accurately approximated when evaluating the retarded field along the worldline.
Following \cite{Poisson} the expansion of the square root of the van Vleck determinant around $\sigma, \sigma^\alpha = 0$ is
\begin{widetext}
\begin{eqnarray}
    \Delta^{1/2} (\bar{z}^\alpha, \bar{z}^{\alpha^\prime} ) = 1 + \frac{1}{12} \, R_{\alpha \beta} ( \bar{z} ) \, \sigma^\alpha \, \sigma^\beta - \frac{1}{36} \, R_{\alpha \beta ; \gamma} \, \sigma^\alpha \, \sigma^\beta \, \sigma^\gamma + \ldots
\label{vanVleck}
\end{eqnarray}
where the $\ldots$ denotes terms of higher order in the expansion. The covariant derivative with respect to the worldline coordinate of the above quantity is
\begin{eqnarray}
    \nabla_{\bar{z}^\nu} \Delta^{1/2} (\bar{z}^\alpha, \bar{z}^{\alpha^\prime} ) = \frac{1}{6} \, R_{\alpha \beta} (\bar{z}) \, \sigma^\alpha _{~ ; \nu} \, \sigma^\beta + \frac{1}{12} \, R_{\alpha \beta ; \nu} (\bar{z}) \, \sigma^\alpha \, \sigma^\beta + \ldots
\end{eqnarray}
Lastly, $\sigma ^\alpha _{~; \nu}$ can be expanded around small $\sigma$ to give
\begin{eqnarray}
    \sigma ^\alpha _{~;\nu} = g^\alpha _{~\nu} - \frac{1}{3} \, R^\alpha _{~\gamma \nu \delta} (\bar{z}) \, \sigma^\gamma \, \sigma ^\delta + \ldots
\end{eqnarray}
so that the gradient of $\Delta ^{1/2}$ becomes
\begin{eqnarray}
    \nabla_{\bar{z}^\nu} \Delta^{1/2} (\bar{z}^\alpha, \bar{z}^{\alpha^\prime} ) =  \frac{1}{6} \ R_{\nu \alpha} (\bar{z}) \, \sigma ^\alpha - \frac{1}{18} \, R ^\lambda _{~\alpha \beta \nu; \lambda} (\bar{z}) \, \sigma^\alpha \, \sigma^\beta  + \ldots
\label{grad_vanVleck}
\end{eqnarray}

Next, we expand these quantities and $g_{ret}^\Lambda$ around $\tau^\prime = \tau+s$ with $s$ such that $\Lambda s \gg 1$. The expansions of some relevant quantities involving the world function $\sigma$ are (see Appendix A for details)
\begin{eqnarray}
    s &=& \tau^\prime - \tau \\
    \sigma (\tau, \tau^\prime) &=& - \frac{1}{2} \, s^2 + {\cal O} (s^4) \\
    \sigma_\mu (\tau, s) &=& - s \: u_\mu (\tau) - \frac{s^2}{2} \: \frac{D u_\mu (\tau)}{d\tau} - \frac{s^3}{6} \: \frac{D^2 u_\mu (\tau)}{d\tau^2} + {\cal O} (s^4)     \label{sigma_mu}
\end{eqnarray}
Using these results, the function $g_{ret}^\Lambda$ is
\begin{eqnarray}
    g_{ret} ^\Lambda = \sqrt{ \frac{ 8}{ \pi} } \: \Lambda^2 \: e^{ - 2 \Lambda^4  \sigma^2 }  = \sqrt{ \frac{ 8}{ \pi} } \: \Lambda^2 \: e^{ - \Lambda^4  s^4 /2 } + {\cal O} (s^6)
\end{eqnarray}
Lastly, we expand $\nabla_{\bar{z}^\nu} g_{ret}^\Lambda$ in powers of $s$
\begin{eqnarray}
    \nabla _{\bar{z} ^\mu} g_{ret}^\Lambda &=& \partial _{\bar{z}^\mu} g_{ret}^\Lambda = \sigma_\mu \left( \frac{ \partial \sigma}{ \partial s} \right)^{\!\!-1} \!\! \frac{\partial g_{ret}^\Lambda}{\partial s} = \left\{ u_\mu (\tau) + \frac{ s}{2} \: \frac{Du_\mu (\tau)}{d\tau} + \frac{ s^2 }{6} \: \frac{D^2u_\mu (\tau)}{d\tau^2} + {\cal O} (s^3) \right\}  \frac{\partial g_{ret}^\Lambda}{\partial s}
\end{eqnarray}
Applying these results to (\ref{starting_pt}) gives
\begin{eqnarray}
    && \left( m_0 - \frac{e^2}{2} \, g_{(1)} (r) + e^2 c_{(0)} (r) - e^2 \int_{\tau_i} ^\tau d\tau^\prime \, V(\bar{z}^\alpha, \bar{z}^{\alpha^\prime} ) \right) \frac{D\bar{u}_\mu}{d\tau} \nonumber \\
    && {\hskip1.5in}  = e \, w_\mu ^{~\nu} [\bar{z}] \, \phi_\nu ^{tail} (\bar{z})  + e^2 \sum_{n=1} ^\infty \left( g_{(n+1)} (r) u_\alpha ^{(n+1)} [\bar{z}] + c_{(n)} (r) v_\alpha ^{(n)} [\bar{z}] \right) 
\label{mean_2}
\end{eqnarray}
with $r = \tau - \tau_i$ being the elapsed proper time since $\tau_i$.

The expressions for the $r$-dependent coefficients ($g_{(n)}$ and $c_{(n)}$) and the trajectory-dependent vectors ($u_\mu ^{(n)}$ and $v_\mu ^{(n)}$) can be found in Appendix B. However, only the relevant factors in the limit of very large $\Lambda$ will be given here
\begin{eqnarray}
    c_{(0)} (r) = - \frac{ \Lambda }{ 2^{1/4} \sqrt{\pi} } \, \gamma \left( \frac{1}{4}, \frac{ r^4 \Lambda^4}{2} \right)  ~~ && ~~ g_{(1)} (r) = - \Lambda \, \frac{ 2^{7/4} }{\sqrt{\pi} } \, \gamma \left( \frac{5}{4}, \frac{r^4 \Lambda ^4}{2} \right) \label{h} \\
    c_{(1)} (r) = \frac{ 1 }{ \sqrt{\pi} } \, \gamma \left( \frac{1}{2}, \frac{ r^4 \Lambda^4}{2} \right) ~~ && ~~ g_{(2)} (r) = \frac{ 2 }{ \sqrt{\pi} } \, \gamma \left( \frac{3}{2} , \frac{ r^4 \Lambda^4}{2} \right) \label{Aone}  \\
    v_\mu ^{(1)} [\bar{z}] &=& \frac{1}{6} w_\mu ^{~\nu} [\bar{z}] \, R_\nu ^{~\alpha} \, \bar{u}_\alpha \\
    u_\mu ^{(1)} [\bar{z}] = \frac{1}{2} \, \bar{a}_\mu ~~ && ~~ u_\mu ^{(2)} [\bar{z}] = \frac{1}{3} \, w_\mu ^{~\nu} [\bar{z}] \, \frac{ D \bar{a}_\nu}{d\tau}
\end{eqnarray}
\end{widetext}
where $\gamma (a,b) = \Gamma(a) - \Gamma(a,b)$ is the incomplete gamma function. The coefficients (\ref{h})-(\ref{Aone}) vary over a time-scale of $\sim \Lambda^{-1}$ after which they are effectively constant. Figure (\ref{fig1}) shows the $r$-dependence of these functions. Note also that, at the initial time, all of these functions vanish. More will be said below concerning these properties and their implication for the validity of the quasi-local expansion.

\subsection{Scalar ALD Equation}

Most of the terms on the right side of (\ref{mean_2}) are irrelevant in the sense that they are inversely proportional to powers of $\Lambda$ so that for time-scales much longer than $\Lambda^{-1}$ these terms can be ignored. In this limit the effect of the high energy physics is ignorable and the effective particle dynamics is described by
\begin{eqnarray}
    && m(\tau; \bar{z} ] \, \frac{D \bar{u}_\mu}{d\tau}  \non \\
    && ~= F_\mu ^{ext} (\tau) + \frac{e^2}{3} \, g_{(2)} (r) \, w_\mu ^{~\nu} [\bar{z}] \, \frac{D \bar{a}_\nu}{d\tau} \non \\
    && {\hskip0.25in} + \frac{ e^2}{ 6} \, c_{(1)} (r) \, w_\mu ^{~\nu} [\bar{z}] \, R_{\nu \alpha} (\bar{z}) \, \bar{u}^\alpha + e \, w_\mu ^{~\nu} [\bar{z}] \, \phi_\nu ^{tail} (\bar{z}) \non \\
    && {\hskip0.25in} + {\cal O} (\Lambda^{-1}) 
\label{eff_mean_eom}
\end{eqnarray}
where the time-dependent and trajectory-dependent effective mass is
\begin{eqnarray}
    && m(\tau; \bar{z}] \non \\
    && ~= m_0 - e^2 \left( \frac{1}{2} \, g_{(1)} (r) - c_{(0)} (r) + \!\! \int _{\tau_i} ^\tau \!\! d\tau^\prime \, V(\bar{z}^\alpha, \bar{z}^{\alpha^\prime} ) \right) \non \\
\end{eqnarray}
and an external force $F_\mu ^{ext}$ responsible for accelerating the charge has been included.
%It is interesting to note that if $\Lambda^{-1}$ is on the order of the classical radius of the particle $e^2/m^2$ then the next-to-leading order terms linear in $\Lambda^{-1}$ may become important. More will be said about this below.

In the limit that $\Lambda \rightarrow \infty$, the terms involving $c_{(0)}(r)$ and $g_{(1)}(r)$
both diverge linearly with $\Lambda$ and the bare mass $m_0$ gets
shifted by an infinite amount $\delta m =
e^2 ( g_{(1)}/2 - c_{(0)} )$. The mass is then
renormalized to $m_{ren} = m_0 - \delta m$, which is a constant
with respect to $\tau$ in the infinite $\Lambda$ limit. If the
regulator is finite but very large then the shift $\delta m$ is
finite and large and the dressed mass has a very weak dependence
on time (see Fig.(\ref{fig1})).

If we now make $\Lambda$ infinite and use the limiting values for $g_{(2)}$ and $c_{(1)}$
\begin{eqnarray}
    && \lim _{\Lambda \rightarrow \infty} g_{(2)} (r) = 1 = \lim _{\Lambda \rightarrow \infty} c_{(1)} (r) 
\end{eqnarray}
for a constant elapsed proper-time $r$ then the low-energy effective particle dynamics is
\begin{eqnarray}
    m_{ren} (\tau; \bar{z}] \, \frac{D \bar{u}_\mu}{d\tau} &=& F_\mu ^{ext} (\tau) + f_\mu [\bar{z}]
\label{ren_mean_eom}
\end{eqnarray}
where the renormalized effective mass is
\begin{eqnarray}
    m_{ren} (\tau; \bar{z}] = m_{ren} - e^2 \int_{\tau_i} ^\tau d\tau^\prime \, V ( \bar{z}^\alpha, \bar{z}^{\alpha^\prime})
\end{eqnarray}
and the self-force is
\begin{eqnarray}
    f_\mu [\bar{z}] = w_\mu ^{~\nu} [\bar{z}] \left( \frac{e^2}{3} \, \frac{D \bar{a}_\nu}{d\tau} + \frac{ e^2}{ 6 }  R_{\nu \alpha} (\bar{z}) \bar{u}^\alpha + e \, \phi_\nu ^{tail} (\bar{z}) \right) \non \\
\label{self_force_1}
\end{eqnarray}
Equations (\ref{ren_mean_eom}) through (\ref{self_force_1}) are the main results of this section. The appearance of the tail term (\ref{tail_term}) implies that the radiation reaction is nonlocal because of its dependence on the past behavior of the classical worldline. This equation for the radiation reaction from scalar charges was first obtained by Quinn \cite{Quinn} based on
earlier work of Quinn and Wald \cite{QW} and Mino, Sasaki and Tanaka
\cite{MST} on gravitational radiation reaction.

Detweiler and Whiting \cite{Detweiler_Whiting} have obtained this
result by decomposing the derivative of the retarded field
$\phi_{ret}$ into a singular piece $\phi^S_\mu$, containing the
diverging contribution that renormalizes the mass, and a regular
piece $\phi^R_\mu$, which contributes to the self-force and is
regular on the worldline. From the previous section we can
construct these quantities within our regularization scheme to
find that for $\Lambda \rightarrow \infty$
\begin{eqnarray}
    \phi^S_\mu &=& - e \sqrt{ \frac{8}{\pi} } \, \Lambda^2 \, \bar{u}_\mu + e \left( \frac{1}{2} \, g_{(1)} (r) - c_{(0)} (r) \right) \frac{D \bar{u}_\mu}{d\tau} \non \\
    \phi^R _\mu &=& -e \, \frac{1 - 6 \xi_R}{12} \, R (\bar{z}) \, \bar{u}_\mu + \frac{e}{3} \, \frac{D \bar{a}_\mu}{d\tau} + \frac{ e}{ 6 } \, R_{\mu \alpha} (\bar{z}) \, \bar{u}^\alpha \non \\
    && + \phi_\mu ^{tail} (\bar{z})
\end{eqnarray}
Notice that the first term of the singular part does not
contribute to renormalizing any physical parameters (at the level
of the equations of motion) since $\phi_\mu ^S$ is projected onto
a direction orthogonal to $\bar{u}_\mu$. Likewise, the first term
of the regular part does not contribute to the self-force on the
particle.

In a flat spacetime, the tail term vanishes since there is no curvature available to focus the radiation emitted in the past onto the particle at the present. Furthermore, (\ref{ren_mean_eom}) reduces to the ALD equation for a scalar field
\begin{eqnarray}
    m_{ren} \, \dot{\bar{u}}_\mu =  F_\mu ^{ext} (\tau) + \frac{e^2}{3} \, w_\mu ^{~\nu} [\bar{z}] \, \ddot{\bar{u}}_\nu
\label{flat_ALD}
\end{eqnarray}
This equation was derived in the open quantum system formalism in \cite{JH1}.

\section{Linearized Fluctuations around the Classical Particle Trajectory} \label{Lin}

We now study the effects of the quantum field fluctuations (as
classical stochastic forces) on the low energy dynamics of the
particle. Instead of working directly with the equations for the
linearized fluctuations (\ref{z_flucs_funcder}), which contain
the singular functional derivatives, we begin with
(\ref{stoch_eom}) and assume that the retarded field has been
regularized by the short-distance regulator $\Lambda$ in a
quasi-local expansion.

\subsection{Stochastic ALD Equation}

For large finite values of $\Lambda$ (\ref{stoch_eom}) becomes
the ALD-Langevin equation
\begin{eqnarray}
    m(\tau; z ] \, \frac{Du_\mu}{d\tau} &=& F_\mu ^{ext} (\tau) + f_\mu (z) +  \eta_\mu (\tau; z]
\label{stoch_ALD}
\end{eqnarray}
where the (regulated) self-force is
\begin{widetext}
\begin{eqnarray}
    f_\mu [z] &=& w_\mu ^{~\nu} [z] \left( \frac{e^2}{3} \, g_{(2)} (r) \, \frac{D a_\nu}{d\tau} + \frac{e^2}{6} \, c_{(1)} (r) \, R_{\nu \alpha } (z) \, u^\alpha + e \, \phi_\nu ^{tail} (z) \right) + {\cal O}(\Lambda^{(-1)})
\label{reg_selfforce}
\end{eqnarray}
Of course, one must remember that these expressions are only valid
up to linear order in the fluctuations $\tilde{z}$ about the mean
worldline $\bar{z}$ since higher orders correspond to quantum
corrections that we have assumed to be negligible. Expanding
(\ref{stoch_ALD}) in orders of the fluctuations, the
time-dependent mass and self-force are given by
\begin{eqnarray}
    m (\tau; z ] &=& m (\tau; \bar{z} ]  - e^2 \int_{\tau_i} ^\tau d\tau^\prime \, \tilde{z} ^{\nu^\prime} \frac{\delta}{\delta \bar{z}^{\nu^\prime} } \int_{\tau_i} ^{\tau} d\tau^{\prime \prime} \, V (\bar{z}^\alpha, \bar{z}^{\alpha^{\prime\prime} } ) + {\cal O}( \tilde{z}^2 ) \\
    f_\mu [z] &=& f_\mu [ \bar{z} ] + \int d\tau^\prime \, \tilde{z}^{\nu ^\prime} \frac{\delta}{\delta \bar{z}^{\nu^\prime}} f_\mu [\bar{z}^\alpha] +{\cal O} (\tilde{z}^2)
\end{eqnarray}

Calculating the functional derivative in the mass equation gives
\begin{eqnarray}
    m (\tau; z ] &=& m (\tau; \bar{z} ] - \frac{e}{2} \, \tilde{z}^\nu \phi_\nu ^{tail} (\bar{z}) - e^2 \int_{\tau_i} ^\tau d\tau^\prime \, \tilde{z}^{\nu^\prime} \nabla_{\bar{z}^{\nu^\prime}} V(\bar{z}^\alpha, \bar{z}^{\alpha^\prime} ) + {\cal O} ( \tilde{z}^2 )
\end{eqnarray}
where the factor of $1/2$ in the second term comes from evaluating a delta function on the upper boundary of the $\tau^\prime$ integral. Notice that the linear terms in $\tilde{z}$ vanish in flat space-time since $V$ is identically zero for a massless scalar field in 3+1 dimensions.

Simplifying the self-force fluctuations is slightly more
involved. The calculation amounts to performing the variational
derivative on $f_\mu$ but keeping in mind to expand out the
covariant derivatives, which depend on the classical worldline coordinates. The
result is
\begin{eqnarray}
    f_\mu [z] = f_\mu [\bar{z}] - \kappa _{\mu\alpha} [\bar{z}] \, \tilde{z}^\alpha - \gamma _{\mu\alpha} [ \bar{z}] \, \dot{\tilde{z}}^\alpha - m_{\mu\alpha} [ \bar{z}] \, \ddot{\tilde{z}}^\alpha + r _{\mu\alpha} [ \bar{z}] \, \dddot{\tilde{z}} ^\alpha + {\cal O} (\Lambda^{-1})
\end{eqnarray}
The time- and trajectory-dependent coefficients are complicated
expressions that are not particularly illuminating. These are recorded in Appendix C.

Combining the linearized mass and self-force into (\ref{stoch_ALD}) and using the fact that $\bar{z}$ satisfies the mean equations of motion (\ref{CGEA_funcvar}) results in
\begin{eqnarray}
   M_{\mu\alpha} [\bar{z}] \, \ddot{\tilde{z}}^\alpha + \Gamma_{\mu\alpha} [\bar{z}] \, \dot{\tilde{z}}^\alpha + K_{\mu\alpha} [\bar{z}] \, \tilde{z}^\alpha - e^2 \, \bar{a}_\mu \int_{\tau_i} ^\tau d\tau^\prime \, \tilde{z}^{\nu^\prime} \nabla_{\bar{z}^{\nu^\prime}} V(\bar{z}^\alpha, \bar{z}^{\alpha^\prime} ) = r_{\mu \alpha} [\bar{z}] \, \dddot{\tilde{z}} ^\alpha + \eta_\mu [\bar{z}] + {\cal O} (\Lambda^{-1} )
\label{particle_perts}
\end{eqnarray}
\end{widetext}
where
\begin{eqnarray}
    K _{\mu\alpha} [\bar{z} ] &=& \kappa_{\mu\alpha} [ \bar{z} ] - m (\tau; \bar{z}] \, \partial _{\bar{z}^\alpha} \Gamma_{\mu \gamma} ^\beta \, \bar{u}_\beta \bar{u}^\gamma - \frac{e}{2} \, \bar{a}_\mu \phi _\alpha ^{tail} (\bar{z}) \non \\
    M_{\mu\alpha} [\bar{z}] &=& g_{\mu\alpha} \, m(\tau; \bar{z}] + m_{\mu\alpha} [ \bar{z}]  
\end{eqnarray}
and
\begin{eqnarray}
    \Gamma_{\mu\alpha} [ \bar{z}] &=& \gamma_{\mu\alpha} [\bar{z}] - 2 m(\tau; \bar{z}] \,  \Gamma_{\mu \left( \alpha \right.}^{\, \beta} \bar{u}_{\left. \beta \right)}
\end{eqnarray}
The dynamical equation (\ref{particle_perts}) for the
fluctuations about the classical particle trajectory is the main
result of this section. This is a linear integro-differential
equation for $\tilde{z}$ with a third derivative term and
contains time-dependent coefficients that depend on the
non-Markovian behavior of the mean trajectory. Furthermore,
because of the integration over past times the last term on the
left side depends on the history of the fluctuations. Notice that
this term vanishes in a {\it flat} background so that the
fluctuations then obey a third-order differential equation, which
is Markovian in the sense that given a mean trajectory
$\bar{z}^\mu$ the fluctuations do not depend on their own past
history. So, given a solution to the mean equation of motion for
the classical worldline one could, in principle, solve for the
fluctuations induced by the quantum field fluctuations.

The notation in (\ref{particle_perts}) has been chosen
suggestively since the left side resembles a damped
simple harmonic oscillator with time-dependent mass, damping
factor, and time- and history-dependent spring constant. Notice also that the ``effective mass" $M_{\mu\alpha} [ \bar{z}]$ is not diagonal implying that the inertia of the fluctuations behaves differently in different directions. This feature is exhibited in all of the other coefficients ($\Gamma_{\mu\alpha}$, $K_{\mu\alpha}$ and $r_{\mu\alpha}$) and suggests that the fluctuations of the trajectory in one direction are linked with the fluctuations in the other space-time directions.  Also, from the expression for the radiation reaction on the fluctuations $r_{\mu\alpha} [\bar{z}]$ it should be noted that this is explicitly independent of the background curvature and effectively projects $\dddot{\tilde{z}}_\nu$ onto a direction orthogonal to the mean 4-velocity $\bar{u}_\nu$.

If the stochastic term $\eta_\mu$ is ignored then
(\ref{particle_perts}) describes the evolution of small
perturbations away from the semiclassical trajectory and so is useful for studying the linear response of the trajectory to small
perturbations away from the mean worldline. In other words,
setting $\eta_\mu$ to zero (\ref{particle_perts}) gives the
linearization of the ALD-Langevin equation around the
semiclassical worldline $\bar{z}$. Generalizing this to the
self-force due to linearized metric perturbations could be useful
for investigating the stability of numerical calculations of the
inspiral of a small mass black hole into a large mass black hole,
for example \cite{Stability}.

%\newpage
\subsection{Memory and Secular Effects}

A particularly interesting feature related to this is the effect of the non-Markovian term appearing in (\ref{particle_perts})
\begin{eqnarray}
    - e^2 \, \bar{a}_\mu \int_{\tau_i} ^\tau d\tau^\prime \, \tilde{z}^{\nu^\prime} \nabla_{\bar{z}^{\nu^\prime}} V(\bar{z}^\alpha, \bar{z}^{\alpha^\prime} )
\end{eqnarray}
If the fluctuations $\tilde{z}$ grow then the
linearization may not be valid for all times and so one should
then include quantum corrections to our stochastic semi-classical
equations if the environment is a quantum field as is assumed
here. The growth of the fluctuations could be significantly influenced
by the past behavior of $\nabla_\nu V$ and the history of the
fluctuations. The behavior of this term on the fluctuations is
not known yet because of the difficulty in solving
integrodifferential equations and in computing $V$ for many
spacetimes. However, a study in de Sitter space (where $V$ is
known \cite{Burko_Harte_Poisson, Poisson}) may provide a simple
arena for understanding the role of the non-Markovian term and
the nature of the solutions to (\ref{particle_perts}).

%As another application of (\ref{particle_perts}) we consider the motion of electrons in strong electromagnetic fields. With the recent advancement of ultraintense lasers (intensity $\sim 10^{22}$ W/cm$^2$) it may be possible to study the effects of strong radiation damping. Many radiation damping studies are carried out for the regime of weak damping, which allows the third-order ALD equation to be reduced to a second-order differential equation by essentially expanding the solution in powers of the particle's classical radius. But in the case of strong external fields, effects of quantum fluctuations and strong damping on the particle trajectory  may become important. (This also provides a way of probing the domain of validity of classical electrodynamics.) If we assume that the particle can be described as a stochastic semi-classical object then we could use the electrodynamics analogue of (\ref{ren_mean_eom}) and (\ref{particle_perts}) to describe the effect of the field fluctuations on the particle's motion. It turns out (see next section) that these effects will probably occur for wavelengths smaller than the Compton wavelength of the particle \cite{Koga}.

Aside from these technical considerations, these equations are
applicable for any type of noise on the particle trajectories. In
many cases, the source of the noise acting on the system of
interest may not be known and is put in by hand in a
phenomenological description of the particle dynamics via the
stochastic equation
\begin{eqnarray}
    m (\tau; z] \, \frac{D u_\mu}{ d\tau} = F_\mu ^{ext} (\tau) + f_\mu (z) + \eta_\mu ^{+}
    \label{addnoise_ALD}
\end{eqnarray}
where the superscript $(+)$ is to remind us that this term was
added in by hand, as opposed to being derived, like our treatment
of $\eta_\mu [\bar{z}]$. This stochastic force could have a
classical origin (e.g. high-temperature thermal fluctuations of surrounding gas) or
it could have no known single identifiable origin. Furthermore,
since the noise $\eta_\mu ^{+}$ is not derived from an initially
closed system it is likely to be inconsistent with the dynamics
of the trajectory. (See, e.g., \cite{HPZ})  In any case, one
would also have to specify the noise correlator $\langle \eta_\mu
^{+} (\tau) \eta_{\mu^\prime} ^{+} (\tau^\prime) \rangle_{\eta^+}$ as
it suits the model.

With this proviso (no guarantee for consistency)  the analysis of
this section carries over.
%, with $\eta_\mu$ replaced by $\eta_\mu ^{+}$.
Given any kind of noise the fluctuations around the mean
trajectory of the particle moving through its own (classical)
field subjected to the self-force from radiation reaction is given
by (\ref{particle_perts}), but with $\eta_\mu [\bar{z}]$ replaced
by $\eta_\mu ^+$
\begin{eqnarray}
   && M_{\mu\alpha} [\bar{z}] \, \ddot{\tilde{z}}^\alpha + \Gamma_{\mu\alpha} [\bar{z}] \, \dot{\tilde{z}}^\alpha \! + K_{\mu\alpha} [\bar{z}] \, \tilde{z}^\alpha \non \\
   && {\hskip0.25in} - e^2 \, \bar{a}_\mu \!\! \int_{\tau_i} ^\tau \!\! d\tau^\prime \, \tilde{z}^{\nu^\prime} \nabla_{\bar{z}^{\nu^\prime}} V(\bar{z}^\alpha, \bar{z}^{\alpha^\prime} ) \non \\
   && {\hskip 0.5in} = r_{\mu \alpha} [\bar{z}] \, \dddot{\tilde{z}} ^\alpha + \eta_\mu ^+ + {\cal O} (\Lambda^{-1} )
\end{eqnarray}

Since in this discussion we don't need to worry about quantum
corrections from higher order loops in the effective action we
could go beyond the linear order in the fluctuations of the
particle trajectory and expand the solutions to
(\ref{addnoise_ALD}) in powers of the  coupling constant $e$ (
denoted in the subscript) $z = z_0 + z_1 + z_2 + \ldots$.
Assuming that $\eta_\mu ^+$ is ${\cal O}(e)$ and depends on $z$
and recalling that $f_\mu$ is quadratic in the coupling we find a
non-local and causal contribution to the total force on the
particle coming from the fluctuations of the stochastic force
(see Appendix D for details)
\begin{eqnarray}
    F^{drift} _\mu = \int d\tau^\prime \, d\tau^{\prime\prime} \, F_\mu ^{\rho\sigma} (\tau, \tau^\prime, \tau^{\prime\prime} ) \, \big\langle \eta_\rho ^+ (z_0 ^{\alpha^\prime}) \, \eta_\sigma ^+ (z_0 ^{\alpha^{\prime\prime}} ) \big\rangle _{\eta^+} \non \\
    \label{drift}
\end{eqnarray}
which is of the same order as the self-force. It seems that the
stochastic noise would cause the particle to drift off from the
background trajectory $z_{(0)}$ determined by the external force. The
deviation could build up over time as indicated by the integral
above. This may have interesting consequences for astrophysical
sources with some stochastic behavior described by a classical
noise. If such physical situations exist, this noise-induced
drift may give rise to a secular effect which could alter the
waveform templates (of events expected to be seen by
gravitational interferometers like LIGO and LISA) calculated
without including such stochastic secular effects.
%However, in flat spacetime one can show that (\ref{drift}) vanishes to all orders of the coupling showing that the overall effect of noise does not alter the expected inertial motion.

A similar expression to (\ref{drift}) can be derived in our open
quantum system framework by writing the effective action to cubic
order in the fluctuations. Doing this reveals a term like
(\ref{drift}) but this requires a much more careful analysis that
goes beyond the stochastic semi-classical approximation adopted
here.

%Another interesting feature to investigate is the possible coherent noise-induced motion (FUTURE PAPER???) (similar, in some sense, to the ponderomotive force in plasmas), which might give detectable signatures to the outgoing radiation. For instance, if we assume that there is a noise source that is rapidly oscillating on a time-scale much shorter than the time-scale of the motion without the rapid oscillations present (the ``slow" motion) then time-averaging over the fast motions can give rise to a drift in the slow motion of the particle. It would be interesting to see the effects of the quantum field fluctuations (in this manner) on the slow motion of the particle under the influence of some external force.

%\newpage
\section{Discussions}

%\subsection{Validity of the Quasi-Local Expansion, Runaway Solutions and the Quantum Regime}
\subsection{The Quantum Regime and the Validity of the Quasi-Local Expansion and Order-Reduction}

The previous sections focused on the Feynman-Vernon formalism and various approximations to obtain the low-energy effective dynamics of the particle, both for its semiclassical (mean) and stochastic semi-classical (mean and stochastic fluctuations) motion. Here, the domain of validity of the quasi-local expansion and this semi-classical treatment will be explored and compared with the relevant scales for weak and strong radiation damping.

%The effective field theory paradigm introduced a regulator $\Lambda$ for controlling the ultraviolet divergences appearing in the direct part of the retarded Green's function such that $\Lambda^2 \sigma \gg 1$ with $\sigma$ small and approaching zero. After expanding $\sigma$ near coincidence the time-scale of the quasi-local expansion $\Delta \tau=s$ followed by requiring

In the effective field theory paradigm a regulator $\Lambda$  is
introduced for controlling the ultraviolet divergences appearing
in the direct part of the retarded Green's function such that
$\Lambda^2 \sigma \gg 1$ with $\sigma$ small and approaching
zero. After expanding $\sigma$ near the coincidence limit the
time-scale of the quasi-local expansion $\Delta \tau=s$ is
governed by
\begin{eqnarray}
    \Delta \tau \gg \Lambda^{-1}
\end{eqnarray}
Recall that for times larger than $\sim \Lambda^{-1}$ the time-dependent coefficients in (\ref{eff_mean_eom}) and (\ref{stoch_ALD}) rapidly approach constant values (see Fig.\ref{fig1}). But there are other scales to consider. We have been working at tree-level in both the particle and the field so that $\Delta \tau$ must be much larger than the scale for 1-loop field and 1-loop particle corrections. This implies that the time interval $\Delta \tau$ should be much longer than the time-scale for creating pairs
\begin{eqnarray}
    \Delta \tau \gg \frac{\hbar}{m} = \lambda_C
\end{eqnarray}
where $\lambda_C$ is the Compton wavelength associated with the scalar particle. The dynamics (\ref{eff_mean_eom}) and (\ref{stoch_ALD}) are therefore valid using this framework provided that
\begin{eqnarray}
    \Delta \tau \gg \lambda_C \gg \Lambda^{-1}
\end{eqnarray}

Nevertheless, the presence of the third $\tau$-derivative in (\ref{eff_mean_eom}),
(\ref{stoch_ALD}) and (\ref{particle_perts}) requires the
specification of the initial position, velocity, and acceleration
to obtain unique solutions. This is problematic, since for a
vanishing external force $F_\mu ^{ext}$ one still requires an
initial acceleration to solve the equations. But, if there is no
force accelerating the charge then what causes the particle to
accelerate? Furthermore, the particle may experience unbounded acceleration so that its kinetic energy increases with time to infinity. Since scalar dynamics is an energy-conserving theory where then could this energy arise? These problems are related to the infinite self-energy of a (classical) charged point particle.

In the {\it classical} theory of scalar fields and particles
these issues can be resolved if one instead gives  the particle a
finite size $r_0$ for which the self-energy is roughly $e^2/r_0$
\cite{LL}. If this energy composes its rest mass then
\begin{eqnarray}
    r_0 \sim \frac{ e^2}{ m}
\end{eqnarray}
and represents the ``size" of the particle. In
\cite{LL}, an approximation, amounting to an asymptotic expansion in
powers of $r_0$, called the Landau approximation (also known as
order-reduction), is employed to obtain solutions that require
only an initial position and velocity and are also free from
run-away solutions. The Landau
approximation converts the ALD equation (of third order) to the
so-called Landau-Lifshitz equation (of second order). We will use
these names to distinguish between these equations.

In order-reduction, the lowest order solution is found by simply ignoring the self-force so that the radiation damping is assumed weak. The time-scale of the dynamics is then determined mostly by the external force so that if $F_\mu ^{ext}$ varies on a scale $\lambda_{ext}$ then $\Delta \tau \sim \lambda_{ext}$. In curved spacetimes the self-force will be weak if $r_0 \ll \Delta \tau$ and the length scale associated with the spacetime curvature $\lambda_R$ is large (i.e. small curvature) so that $\lambda_R \gg r_0$. It then follows that for weak damping in the semi-classical domain that the quasi-local expansion and the Landau approximation are valid  provided that
\begin{eqnarray}
    \Delta \tau \sim \lambda_{ext} \gg r_0, \lambda_C \gg \Lambda^{-1} {\rm ~~ and ~~ } \lambda_R \gg r_0
\end{eqnarray}

Recently, in the context of plasma physics, \cite{Koga}
has investigated the validity of the Landau approximation for the
classical ALD equation by numerically integrating the
Landau-Lifshitz equation forward in time and, using the final
position, velocity, and acceleration from that, integrating the
ALD equation backward in time. If the initial position and
velocity of the particle differ significantly from the
backward-evolved solution of the ALD equation at the initial time
then one can assume the Landau approximation has broken down.
Koga does this for a counter-propagating electron and
ultraintense laser beam (intensity $\sim 10^{22}$ W/cm$^2$). He
finds that the Landau approximation is valid so long as the laser
wavelength $\lambda_0$ is greater than the Compton wavelength.
For $\lambda_0$ much smaller than $\lambda_C$, he finds
disagreement between the solutions of the Landau-Lifshitz and ALD
equations. However, these equations cannot be fully trusted since
quantum effects should become important. In this domain, while we cannot directly apply our results to this problem we can use the closely related closed-time-path (CTP) formalism to incorporate the effects of the quantum loop corrections to the (quantum) particle dynamics.

\subsection{Decoherence}

This brings us to an important issue, namely, how strongly must
the particle worldline decohere in order for a stochastic
semi-classical treatment to be applicable? A measure of the
decoherence of a system is the {\it decoherence functional}
\cite{conhis} $D[ \alpha(z), \alpha (z^\prime) ]$ where
$\alpha(z)$ specifies a particular coarse-grained history of the
worldline. If we take the history to be %fine-grained so that
$\alpha(z) = z$ then the decoherence functional is
\begin{eqnarray}
    D[ \alpha(z), \alpha(z^\prime) ] &=& D[z,z^\prime] \\
        &=& \rho_S (z_i, z_i ^\prime ) \, e^{ \frac{ i}{\hbar} S_{CGEA} [z,z^\prime] }
\end{eqnarray}
and is closely related to the CTP effective action
\cite{Calzetta_Hu}. In fact, the norm of $D$ is proportional to
the norm of the influence functional. The system is said to
decohere if $|D|$ approaches zero. We showed earlier
that the CGEA is proportional to $z^-$ and so there is minimal
decoherence for two nearby histories. This implies that
coarse-graining the quantum field must be complemented by an
additional coarse-graining of the worldline in order for the
particle to be sufficiently decohered. If the mechanism for
decoherence is efficient then this additional coarse-graining
should be minimal and the stochastic semi-classical description
that we have used in this paper is applicable. For our problem,
recall that the CGEA contains an imaginiary part that is
proportional to $j^- \cdot G_H \cdot j^-$ so that the norm of $D$
is
\begin{eqnarray}
    | D[z,z^\prime] | &=& e^{ - \frac{1}{4\hbar} j^- \cdot (16 \pi^2 G_H ) \cdot j^-}
\end{eqnarray}
showing that the quantum field fluctuations in the environment is
the mechanism for decoherence. It is therefore necessary to
understand the behavior of the Hadamard function to show that
there exists a well-defined stochastic semi-classical regime for
the particle. This requires knowing the field modes on the background
space-time, which is a significant problem in its own right. For
simple scenarios the decoherence functional can be calculated and
the decoherence time approximated. But, in general, one needs to
determine if decoherence is fast enough on a case by case basis.

\subsection{Problems with putting noise in by hand}

Towards the end of Sec. 5 we replaced the classical stochastic manifestation of the quantum fluctuations by a noise source $\eta_\mu ^+$ that has some specified noise correlator.
Aside from the issue about keeping only to tree-level in the particle and field variables, how is using $\eta_\mu ^+$ different from all of the sophisticated machinery used in the previous sections?

%The first point to make is that the noise put in by hand
%$\eta_\mu ^{+}$ is only time-dependent and not trajectory
%dependent as it is in (\ref{stoch_eom}) where the noise is
%obtained from coarse-graining the quantum field fluctuations. The
%noise $\eta_\mu ^{cl}$ appears as an external force ignorant of
%what the particle is doing and how it is moving. Using the
%Feynman-Vernon approach, the noise $\eta_\mu (\tau; z]$ is
%sensitive to the particle's trajectory. Because this stochastic
%force is related to the quantum field fluctuations, which may
%have different behavior in regions of a curved spacetime (e.g.
%near a Schwarzschild horizon as compared with far away in the
%asymptotically flat region), the effects on the particle's motion
%may be quite different depending on where the particle is. The
%fundamental difference is that in our approach the noise in the
%environment (field) is self-consistently determined with the
%system (particle) dynamics.
%This feature shows some of the self-consistency of the Feynman-Vernon
%approach by incorporating the behavior of the environment into the system dynamics.

Unlike in our approach, inserting a source of noise by hand implies that it can have any
correlator one wants. The stochastic two-point function of $\eta_\mu ^{+}$
\begin{eqnarray}
    \big\langle \{ \eta_\mu ^{+} (\tau), \eta_\nu ^{+} (\tau^\prime) \} \big\rangle _{\eta^{+}} = N_{\mu \nu} (\tau, \tau^\prime) ~,
\label{fudged_noise}
\end{eqnarray}
otherwise known as the noise kernel, can be chosen at will. Some physical reasoning, for instance, a high temperature environment, might suggest that $N_{\mu\nu} (\tau, \tau^\prime) \sim \delta (\tau - \tau^\prime)$, i.e. white noise. While white noise certainly simplifies the calculations, there is little justification for simply stating the form of the noise kernel and expecting the dynamics of the system to be consistent with that source. Using the Feynman-Vernon formalism, coarse-graining over the quantum environment naturally gives rise to the noise kernel appropriate for the stochastic force fluctuations
\begin{eqnarray}
    && \big\langle \{ \eta_\mu [\bar{z}^\alpha] , \eta_\nu [\bar{z}^{\alpha^\prime} ] \} \big\rangle _{\xi} \non \\
    && {\hskip0.25in} = \hbar \, e^2 \, \vec{w}_{\left( \mu \right.} [\bar{z}^\alpha] \vec{w}_{\left. \nu \right)} [\bar{z}^{\alpha^\prime}] \, G_H (\bar{z}^\alpha, \bar{z}^{\alpha^\prime} ) 
\label{noise_corr}
\end{eqnarray}
This coarse-graining ensures that the dynamics of the system
evolves consistently with that of the environment. This is an
important statement that cannot be overstated because it is this
feature which gives rise to consistent fluctuation-dissipation
relations (FDR). Inserting noise by hand most likely violates
such a relation.
% such a thing because of the arbitrariness of the noise kernel.

While FDR's won't be discussed here in great detail, \cite{RHA}
have studied the FDR's for $n$ particles interacting with a
quantum field for certain trajectories (e.g. uniformly
accelerated). They have shown the existence and self-consistency
of these relations and further introduce correlation-propagation
relations relating the correlations of particles to each other
through the (causal) propagation of the quantum field. These
correlation-propagation relations can only be formed if the
particles interact directly through causal influences. The open
quantum system viewpoint, therefore, shows the interrelated
influences of the system and environment through these types of
self-consistent relations. This is an important feature lacking
in a model that has noise put in by hand.

Another problem with adding noise in by hand is that the noise can be a significant contribution to the quantum two-point function of the system if one is interested in the quantum particle behavior. So, having the correct form for the noise kernel is important for getting the correct expression for the quantum two-point functions.

It has been shown in \cite{Calzetta_Roura_Verdaguer},  using the CTP generating functional for quantum correlation functions of the system variables, for quantum Brownian motion as well as for stochastic semi-classical gravity (see next section), that certain quantum two-point functions are related to stochastic two-point functions. As an example, the symmetrized quantum two-point function of a Brownian particle's position can be written as
\begin{eqnarray}
    \frac{1}{2} \, \big\langle \{ \hat{x} (t), \hat{x} (t^\prime) \} \big\rangle = \big\langle \, \big\langle X(t) X(t^\prime) \big\rangle _\xi \, \big\rangle_{X_i, p_i}
\end{eqnarray}
where $X$ is a solution of the QBM Langevin equation $L \cdot X =
\xi$ for the appropriate linear operator $L(t,t^\prime)$, and
$\langle \ldots \rangle _{X_i, p_i}$ denotes the average over all
possible initial positions and momenta with respect to the
reduced Wigner function for the initial state of the particle.
The solutions of the QBM Langevin equation consist of a
homogeneous part, containing all the information about the
initial conditions of the system, and a term describing effects
due to the interactions between the particle and oscillators.
\begin{widetext}
\begin{eqnarray}
    X(t) = X_0 (t) + \int dt^\prime \, G_{ret} (t, t^\prime) \xi (t^\prime)
\end{eqnarray}
The symmetrized quantum two-point function becomes
\begin{eqnarray}
    \frac{1}{2} \, \big\langle \{ \hat{x} (t), \hat{x} (t^\prime) \} \big\rangle = \big\langle X_0 (t) X_0 (t^\prime) \big\rangle _{X_i, p_i} + \int dt_1 \int dt_2 \, G_{ret} (t, t_1) N(t_1, t_2) G_{ret} (t_2, t^\prime)
\end{eqnarray}
\end{widetext}
where $N = \langle \xi (t_1) \xi(t_2) \rangle _\xi$ is the noise kernel, given by the stochastic correlator of $\xi$. The first term involving the homogeneous solution of the Langevin equation represents the dispersion in the initial conditions and is called the {\it intrinsic fluctuations}. The second term involving the noise kernel, and hence the quantum fluctuations of the environment, represents the correlations with the environment through the stochastic fluctuations and is called the {\it induced fluctuations}. It turns out that if the homogeneous solution $X_0$ decays exponentially fast then for late times, at least times larger than the decay time, the quantum two-point function is determined entirely by the noise kernel, that is, the induced fluctuations. And so, in this sense, all of the information about the quantum correlations of the system degrees of freedom is encoded in the stochastic correlations. This shows another drawback to putting noise in by hand. The induced fluctuations contain information about the quantum fluctuations of the system variables. But this will not be true with some arbitrary noise kernel chosen at will.

%For the case of radiation reaction, the quantum two-point
%functions of $\hat{z}_\mu$ can be shown to be related to the
%stochastic two-point functions of $\eta_\mu$, at least for
%non-relativistic motion in the frame determined by the
%instantaneous initial data and after the Landau approximation has
%been done so that the effects of radiation damping are analogous
%to the damping in the QBM model above. Complications due to the
%gauge freedom of the worldline parametrization  as well as the
%nonlinear term $\sqrt{- \dot{z}_\mu \dot{z}^\mu }$ appearing in
%the system-environment interaction make an explicit association
%between the quantum and stochastic two-point functions difficult
%to do. However, the qualitative statement that the induced
%fluctuations give an important contribution to the quantum
%two-point function is still true.

\subsection{Similarities with Stochastic Semiclassical Gravity}

The features of the particle dynamics seen in the above discussions are typical of nonequilibrium open quantum systems. History-dependent behavior is present in the equations of motion for the system and if a renormalization procedure is required it is usually a time-dependent prescription, as seen earlier with the renormalized mass $m_{ren}(\tau)$. Furthermore, the noise correlator (\ref{noise_corr}) is generically non-local in time and is determined by the quantum fluctuations of the environment variables. This formalism does not allow for arbitrary noise kernels since this would destroy the self-consistency between the system and environment evolution. A particular example that contains these features is stochastic semiclassical gravity, which we will briefly describe and compare with below.

Stochastic semiclassical gravity (SSG) is a self-consistent
theory of the stochastic dynamics of a classical spacetime
containing quantum matter fields. SSG goes beyond semiclassical
gravity, for which the geometry is driven by the expectation of
the (renormalized) stress tensor, in that the quantum field
fluctuations also contribute to the spacetime dynamics through a
classical stochastic source. The spacetime is therefore driven by
both the quantum expectation value of the renormalized stress
tensor and a classical stochastic stress-tensor-like object,
$\xi_{ab}$. For an introduction and review of this subject see
\cite{StoGraRev} and \cite{Hu_Roura_Verdaguer}  for a discussion
of the domain of validity of SSG.

As an open quantum system, the quantum field fluctuations are coarse-grained using the CTP formalism of Schwinger and Keldysh (SK) to study the self-consistent evolution of the (classical) geometry. The quantum fluctuations manifest themselves as stochastic noise thereby imparting a stochastic nature to the spacetime. The resulting Einstein-Langevin equation for the linearized metric perturubations $h_{ab}$ is
\begin{eqnarray}
    G_{ab} ^{(1)} [g+h] = \kappa \, \big\langle \, \hat{T}_{ab} ^{(1)} [g+h] \, \big\rangle _{ren} + \kappa \, \xi_{ab} [g]
\label{EinLang}
\end{eqnarray}
The superscript $^{(1)}$ denotes that those quantities contain
all terms to first order in the metric fluctuations $h_{ab}$. It
should be noted that the finite parts of the counterterms needed
to cancel the divergences coming from the stress tensor
expectation value have been absorbed into the definition of
$\langle \hat{T}_{ab} ^{(1)} \rangle_{ren}$. The renormalized
stress tensor expectation value (evaluated in a Gaussian state)
contains an integration over the past history of the metric
fluctuations and so the dynamics is generally non-Markovian. This
is like what is seen in the ALD-Langevin equation
(\ref{stoch_ALD}) where the tail term $\phi_\nu ^{tail} (z)$ is
analogous to the expectation value of the renormalized stress
tensor in (\ref{EinLang}). The (covariantly conserved) stochastic
source tensor $\xi_{ab}$ has zero mean and its correlator is
given in terms of the Hadamard function of the stress tensor
fluctuations $\hat{t}_{ab} = \hat{T}_{ab} - \langle \hat{T}_{ab}
\rangle$
\begin{eqnarray}
    \big\langle \{ \xi_{ab} (x; g], \xi_{cd} (x^\prime; g] \} \big\rangle _\xi = \hbar \, \big\langle \{ \hat{t}_{ab} (x; g], \hat{t}_{cd} (x^\prime; g] \} \big\rangle \non \\
\end{eqnarray}
The correlator of the stress tensor fluctuations on the right side does not vanish on a spatial hypersurface. This reflects the fact that the quantum field correlations are themselves non-local. Compare this with the correlator in (\ref{xi_correlator}) which is also non-local.

SSG also suffers from runaway solutions since the finite contributions to the counterterms needed to cancel the divergences appearing from the expectation value of the stress tensor are quadratic in the curvature. This makes SSG a theory with derivatives higher than two, similar to the ALD type equations derived above, which were of third order in the $\tau$ derivatives. A version of the Landau approximation can be used to reduce the order of the Einstein-Langevin equation to two thereby yielding well-behaved solutions free of the pathologies typical of higher-order derivative theories. Of course, one needs to be careful to use the Landau approximation at scales that are consistent with the derivation of the Einstein-Langevin equation.

Finally, the symmetrized quantum two-point functions of the
metric fluctuations $h_{ab}$ can be written in terms of intrinsic
fluctuations, representing the dispersion in the initial
conditions, and induced fluctuations, encoding the information
about the fluctuations of the quantum matter \cite{Hu_Roura_Verdaguer}. Just like with the particle motion, one cannot simply use any noise kernel for modeling stochastic metric fluctuations.
One needs to do a careful analysis that ensures the self-consistency of the metric and quantum matter dynamics and the existence of fluctuation-dissipation relations.

\section{Summary}

In this paper we have derived the scalar ALD equation for the
quantum expectation value of the worldline for a scalar charged
point particle interacting with its own quantum field as it moves
in a curved spacetime. Our equation (\ref{ren_mean_eom}) for the
low-energy effective particle dynamics agrees with the results
obtained earlier by \cite{Quinn, Detweiler_Whiting, Poisson,
Burko_Harte_Poisson}. If the quantum fluctuations in the field
strongly decohere the worldline then we can ignore the particle's
quantum fluctuations and obtain the semi-classical motion
(\ref{ren_mean_eom}). Invoking an effective field theory point of
view, the singular behavior of the field's retarded Green's
function can be regulated. For sufficiently short times $\Delta
\tau=s$ but still long enough $\Delta \tau \gg \Lambda^{-1}$
compared with the inverse cutoff frequency $\Lambda$ a
quasi-local expansion can be used to obtain the contributions
relevant to the self-force and those that are irrelevant in the
infinite $\Lambda$ limit. This renormalizes the mass of the
particle and shows explicitly the appearance of the expected
$Da_\mu/d\tau$ term that is characteristic of radiation reaction.

While the time-dependent coefficients (\ref{h})-(\ref{Aone})
appearing in (\ref{eff_mean_eom}) seem to suggest that only the
initial position and velocity of the particle are needed, and
hence that a resolution of the problems of pre-acceleration and
run-away solutions has been reached, one should keep in mind that
the quasi-local approximation breaks down for short time
intervals, which includes the instant at the initial time. A more
careful analysis would need to include a more physical initial
state than the factorized one used here.

Fluctuations in the quantum field is expected to affect the
particle's motion causing it to fluctuate by an amount $\tilde{z}$
around the mean trajectory $\bar{z}$ given by solutions to the
semi-classical equation (\ref{eff_mean_eom}). We derived such a
stochastic force $\eta_\mu[\bar{z}]$ with correlators from the
fluctuations of the quantum field and a scalar ALD-Langevin
equation (\ref{stoch_ALD}). The dynamics of $\tilde{z}$ (\ref{particle_perts})
contains a non-Markovian contribution through the past history of
the particle fluctuations. Depending on the behavior of this term
the fluctuations might grow to be large indicating a breakdown of
the linear approximation. In that case it requires the inclusion
of quantum corrections in order to follow the nonlinear evolution
of $\tilde{z}$. On the other hand, ignoring the noise altogether
in (\ref{particle_perts}), one can test the
stability of numerical simulations in inspiral studies,  for
example \cite{Stability}.

Instead of the noise $\eta_\mu$ derived here from fluctuations of
quantum fields one can replace it with some other classical noise
$\eta_\mu ^+$ suitably chosen to model some stochastic source in a
phenomenological description. We can still use
(\ref{particle_perts}) to study the effect of such noises on the
particle trajectory fluctuations. However, since the origin for
the noise is no longer due to a quantum field we need not worry
about keeping up to linear order in $\tilde{z}$ in the
ALD-Langevin equation. Instead, expanding the solution for small
coupling constant $e$ and taking the stochastic expectation value
shows that there is, in general, a non-vanishing force (\ref{drift}) coming completely from the
correlations of the stochastic force.
%\begin{eqnarray}
%\frac{1}{m^2} \, \partial_{z_0 ^\nu} \int d\tau^\prime \, g_{ret} (\tau, \tau^\prime) \, \big\langle \eta_\mu ^+ (z_0 ^\alpha) \,\eta^{+ \nu} (z_0 ^{\alpha ^\prime} ) \big\rangle _{\eta^+}\end{eqnarray}
Along with the self-force, this noise-induced term would cause
the particle to drift off of its background trajectory determined
by the external force (or off of its geodesic motion if $F_\mu
^{ext}$ vanishes). We hope to explore the consequences of this
noise-induced drift in an astrophysical setting  and find
observable effects on the waveforms of the radiation emitted by
the particle and detected by gravitational interferometers like
LIGO and LISA.

%Putting noise in by hand, while perhaps something reasonable to
%do from a practical stand point, should come with some
%reservation. In doing this, a fluctuation-dissipation relation is
%not necessarily guaranteed since a noise kernel for the
%stochastic fluctuations can be chosen at will, although one
%usually takes into account some physical argument (e.g.
%high-temperature thermal fluctuations) to justify a particular
%noise kernel. Furthermore, the stochastic particle fluctuations
%contain some information about the quantum fluctuations of the
%particle, so one could not expect to extract any information
%about the quantum correlation functions of the particle when
%using a noise kernel that is not appropriate for the environment.

In Paper II we will apply the same techniques here to study the
self-forces and the stochastic semi-classical motions of electric
charges coupled to an electromagnetic field and of small black
holes coupled to the background spacetime of a massive black hole,
respectively. We hope to apply these results to more physical
situations such as the motion of charges in strong external fields
and gravitational radiation reaction. %\\

%\noindent {\bf Acknowledgements} 
\begin{acknowledgments}
CRG thanks Phil Johnson for help
in learning the worldline influence functional approach approach and Albert Roura for suggestions in simplifying some derivations in Appendix A. We both
thank Phil Johnson and Albert Roura for interesting discussions
on general issues of radiation reaction. This work is supported
in part by NSF grant PHY03-00710.
\end{acknowledgments}

%\newpage
\appendix

\section{Geodesic Coordinates and the Quasi-Local Expansion}

In this Appendix we derive the quasi-local expansion of $\sigma^\alpha$ in (\ref{sigma_mu}) and some of the relations appearing in Section IV for computing the quasi-local expansion of the regulated direct part of the retarded Green's function. 

\begin{figure}
	\includegraphics[height=6cm]{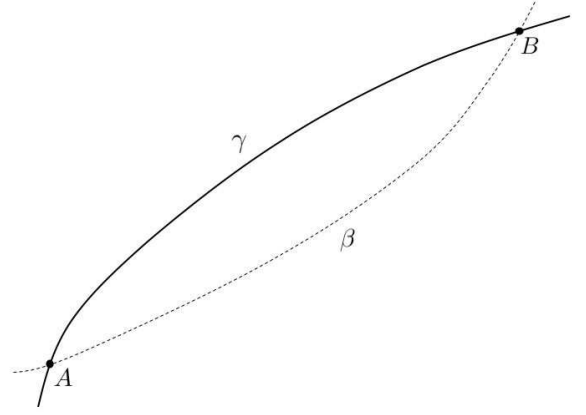}
    	\caption{For two points on a worldline $\gamma$ one can construct a unique time-like geodesic $\beta$ connecting them.}
    \label{fig2}
\end{figure}

We begin by invoking the so-called geodesic coordinates along and near the worldline \cite{LL, Rashevskii}. These coordinates will be denoted with a hat. In geodesic coordinates the metric tensor and the connection coefficients are constructed to vanish along the entirety of the worldline
\begin{eqnarray}
	\hat{g}_{\mu \nu } |_\gamma = 0 ~~~~~ \hat{\Gamma}^\lambda _{\mu \nu } | _\gamma = 0
\end{eqnarray}
so that a vector field on the worldline behaves like a vector field in flat spacetime. The displacement from $A$ to $B$ in Fig(\ref{fig2}) can be written as
\begin{eqnarray}
	\Delta \hat{z}^\alpha \bydefn \hat{z}^\alpha - \hat{z}^{\alpha^\prime} 
\end{eqnarray}
Using a Taylor series to express $\hat{z}^{\alpha^\prime}$ in terms of $\hat{z}^\alpha$ and its derivatives results in
\begin{eqnarray}
	\Delta \hat{z}^\alpha &=& - s \, \hat{u}^\alpha - \frac{s^2}{2!} \, \frac{d \hat{u}^\alpha }{ d\tau} - \frac{s^3}{6!} \, \frac{d^2 \hat{u}^\alpha }{ d\tau^2 } - \ldots \non \\
	&=& - s \, \hat{u}^\alpha - \frac{s^2}{2!} \, \frac{D \hat{u}^\alpha }{ d\tau} - \frac{s^3}{6!} \, \frac{D^2 \hat{u}^\alpha }{ d\tau^2 } - \ldots
\end{eqnarray}
where, in the last line, we have used the fact that the components of the connection vanish in this coordinate system.

Flatness along the worldline implies that the tangent spaces at each point on $\gamma$ can be identified with each other so that tensor manipulations along $\gamma$ can be done in a single tangent space. This implies that $\Delta \hat{z}^{\alpha}$ is a vector in the tangent space that we can interpret as a displacement vector at $B$. Transforming back to the original coordinates gives
\begin{eqnarray}
	\Delta z^\alpha &=& \frac{ \partial x^\alpha }{ \partial \hat{x}^{ \beta} } \, \Delta \hat{z}^{\beta} \non \\
	&=& - s \, u^\alpha - \frac{s^2}{2!} \, a^\alpha - \frac{s^3}{6!} \, \frac{D a^\alpha }{ d\tau } - \ldots
\end{eqnarray}

Using geodesic coordinates appropriate for the geodesic connecting $A$ and $B$ (denoted with a bar instead of a hat) one finds that
\begin{eqnarray}
	\bar{\sigma}^\alpha = (\lambda - \lambda^\prime) \, \bar{t}^\alpha = \bar{y}^\alpha - \bar{y}^{\alpha^\prime} 
\end{eqnarray}
which represents the coordinate difference between $A$ and $B$ in the tangent space at $B$. Here we take $\bar{t}^\alpha$ to be the tangent vector to the geodesic $\beta$ connecting $A$ and $B$ with affine parameter $\lambda^\dprime$ and with coordinates $y$ such that $y |_A = y(\lambda^\prime)$ and $y |_B = y (\lambda)$.  In the original coordinates
\begin{eqnarray}
	\sigma^\alpha = \frac{\partial x^\alpha}{ \partial \bar{x}^\beta } \, \bar{\sigma}^\beta
\end{eqnarray}
Since $\Delta \hat{z}^\alpha$ and $\bar{\sigma}^\alpha$ both represent the (coordinate) difference between $A$ and $B$ in the same tangent space then they are related by the coordinate transformation from the hat to the bar coordinates
\begin{eqnarray}
	\bar{\sigma}^{\beta} = \frac{\partial \bar{x}^{\beta} }{ \partial \hat{x}^{\gamma} } \, \Delta \hat{z}^{\gamma}
\end{eqnarray}
This immediately implies that
\begin{eqnarray}
	\sigma^\alpha (z^\alpha, z^{\alpha^\prime}) &=& - s \, u^\alpha - \frac{s^2}{2!} \, a^\alpha - \frac{s^3}{6!} \, \frac{D a^\alpha }{ d\tau } - \ldots \non \\
	&=& - \sum_{n=1} ^\infty \frac{s^n}{n!} \, \frac{D^n}{d\tau^n} \, u^\alpha(\tau)
\end{eqnarray}
%Since $\sigma^\alpha$ is tangent to the geodesic connecting $A$ and $P$ on the worldline then for a point $x^\prime$ on $\gamma$ in the (past) normal convex neighborhood of $x$ on the worldline, $\Delta^\alpha (x,x ^\prime)$ is the vector at $x$ that generates the unique geodesic connecting those two points. 
There is an interesting interpretation one may attach to this result. Pick a point $x^\prime$ in the normal convex neighborhood of $x$ in some spacetime. These points are connected by a unique geodesic described by $\sigma^\alpha$. If these two points intersect any worldline describing the motion of some particle then in terms of the particle's velocity and its derivatives $\sigma^\alpha = \Delta z^\alpha$. This equality therefore describes a (many-to-one) mapping between particle motions and geodesics.

From the identity $2 \sigma = \sigma_\alpha \sigma^\alpha$ it is possible to show that
\begin{eqnarray}
	\sigma &=& - \frac{s^2}{2} - \frac{ s^4 }{24} + {\cal O} (s^5)
\end{eqnarray}
This also provides a relation between the parametrization of the geodesic and the proper time of the worldline through
\begin{eqnarray}
	( \lambda - \lambda^\prime)^2 = s^2 + \frac{s^4}{12} + {\cal O} (s^5)
\end{eqnarray}
since $2 \sigma = - (\lambda - \lambda^\prime)^2$ is the amount of elapsed proper time along the geodesic. 

Using these results, the covariant derivative of a scalar function, say $g^\Lambda _{ret} (\sigma)$, is
\begin{eqnarray}
	\nabla_\alpha g^\Lambda _{ret} (\sigma) &=& \sigma_\alpha \left( \frac{ \partial \sigma}{ \partial s} \right)^{-1} \frac{ \partial g^\Lambda _{ret} (\sigma) }{ \partial s }  \non \\
	&=& - \frac{\sigma_\alpha}{s} \, \frac{\partial g_{ret} ^\Lambda}{\partial s} + {\cal O}(s^4)
\end{eqnarray}

%\begin{widetext}
\section{Coefficients and Vectors in Regulated Dynamics}

The $r$-dependent coefficients ($r = \tau- \tau_i$) appearing in the equation for the regulated semi-classical dynamics (\ref{mean_2}) are
\begin{eqnarray}
    c_{(n)} (r) &=& - \int_0 ^r ds \, \frac{(-s)^n}{n!} \, g_{ret}^\Lambda (s) \non \\
    &=& ( -1)^{n+1} \frac{ 2^{(n-1)/4} }{ \pi^{1/2} \, n! } \, \Lambda^{1-n} \, \gamma \left( \frac{1+n}{4} , \frac{r^4 \Lambda^4}{ 2} \right) \non \\
\end{eqnarray}
\begin{eqnarray}
    g_{(n)} (r) &=& - \int _0 ^r ds \: \frac{ (-s)^n}{ n! } \: \frac{\partial}{\partial s} \: g_{ret} ^\Lambda (s) \non \\
    &=& (-1)^n \: \frac{ 2^{(n+6)/4} }{ \pi ^{1/2} \: n! } \: \Lambda^{2-n} \, \gamma \left( 1 + \frac{n}{4} , \frac{r^4 \Lambda^4}{ 2} \right) \non \\
\end{eqnarray}
The corresponding worldline-dependent vectors are
\begin{eqnarray}
    u_\mu ^{(0)} &=& 0 \\
    u_\mu ^{(1)} &=& \frac{1}{2} \, \bar{a}_\mu \\
    u _\mu ^{(2)} &=& \frac{1}{3} \, w_\mu ^{~\alpha} \frac{D \bar{a}_\alpha }{d\tau} \\
    u_\mu ^{(3)} &=& \frac{1}{4} \, w_\mu ^{~\alpha} \frac{D^2 \bar{a}_\alpha}{d\tau^2} + \frac{1}{4} \, a_\mu R_{\alpha \beta} u^\alpha u^\beta \\
	&\vdots& \nonumber 
\end{eqnarray}
\begin{eqnarray}
	v_\mu ^{(0)} &=& - \bar{a}_\mu \\
	v_\mu ^{(1)} &=& \frac{1}{6} \, w_\mu ^{~\alpha} R_{\alpha\beta} \bar{u} ^\beta \\
	v_\mu ^{(2)} &=& \frac{1}{6} \, w_\mu ^{~\alpha} R_{\alpha \beta} \bar{a} ^\beta - \frac{1}{6} \, \bar{a}_\mu R_{\alpha \beta} \bar{u} ^\alpha \bar{u} ^\beta \non \\
	&& ~ + \frac{1}{9} \, R^{~~~~~\gamma} _{\mu \alpha \beta ~~; \gamma} \bar{u}^\alpha \bar{u}^\beta \\
	&\vdots& \nonumber 
\end{eqnarray}
For finite and large $\Lambda$ the first-order correction to the self-force $f_\alpha[\bar{z}]$ is 
\begin{eqnarray}
	e^2 \left( g_{(3)}(r) \, u_\alpha ^{(3)} [\bar{z}] + c_{(2)}(r) \, v_\alpha ^{(2)} [\bar{z}] \right)
\end{eqnarray}
which in flat spacetime reduces to $\frac{e^2}{4} \, g_{(3)} (r) \, w_\alpha ^{~\beta} [\bar{z}] \frac{D^2 \bar{a}_\beta}{d\tau^2}$.

\begin{widetext}
\section{Coefficients in Linearized Stochastic Dynamics}

The time- and mean worldline-dependent coefficients appearing in (\ref{particle_perts}) as a result of linearizing the self-force $f_\mu$ (\ref{reg_selfforce}) are
\begin{eqnarray}
    \kappa_{\mu \alpha} (\tau; \bar{z}] &=& \frac{e^2}{3} \, g_{(2)}(r) \left( \frac{d}{d\tau} \left( \partial_{\bar{z}^\alpha} \Gamma_{\beta \nu} ^{\,\gamma} \bar{u}^\beta \bar{u}_\gamma \right) + \partial_{\bar{z}^\alpha} \Gamma_{\beta \nu}^{\, \gamma} \bar{u}^\beta \bar{a}_\gamma - \bar{u}^\beta \Gamma_{\beta \nu}^{\, \gamma} \partial_{\bar{z}^\alpha} \Gamma_{\gamma \delta}^{\, \epsilon} \bar{u}^\delta \bar{u}_\epsilon \right)   \nonumber \\
    && - w_\mu ^{~\nu} [\bar{z}] \left( \frac{e^2}{6} \, c_{(1)}(r) \bar{u}^\beta \nabla_{\bar{z}^\alpha} R_{\nu \beta} + e \nabla_{\bar{z}^\alpha} \phi_\nu ^{tail} (\bar{z}) \right)
\end{eqnarray}
\begin{eqnarray}
    \gamma_{\mu \alpha} (\tau; \bar{z}] &=& \frac{e^2}{3} \, g_{(2)}(r) w_\mu ^{~\nu} [\bar{z}] \left[ 2 \frac{d}{d\tau} \left( \Gamma_{\beta \nu}^{\,\gamma} \bar{u}^{\left( \beta \right.} \right) g_{\left. \gamma \right) \alpha} + \partial_{\bar{z}^\alpha} \Gamma_{\beta \nu}^{\, \gamma} \bar{u}^\beta \bar{u}^\gamma + \Gamma_{\alpha \nu}^{\, \gamma} \bar{a}_\gamma - 2 \bar{u}^\beta \Gamma_{\beta \nu}^{\, \gamma} \Gamma_{\gamma \delta}^{\, \epsilon} \bar{u}^{\left( \delta \right. } g_{\left. \epsilon \right) \alpha} \right] \nonumber \\
    && - \frac{ e^2}{6} \, w_\mu ^{~\nu} [\bar{z}] c_{(1)}(r) R_{\nu \alpha} - 2 \bar{u}^{\left( \nu \right.} g_{\left. \mu \right) \alpha} \left( \frac{e^2}{3} \, g_{(2)}(r) \frac{D \bar{a}_\nu}{d\tau} + \frac{e^2}{6} \, c_{(1)}(r) R_{\nu \beta} \bar{u}^\beta + e \phi_\nu ^{tail} (\bar{z}) \right)
\end{eqnarray}
\begin{eqnarray}
    m_{\mu \alpha} (\tau; \bar{z}] &=& \frac{e^2}{3} \, g_{(2)}(r) \, w_\mu ^{~\nu} [\bar{z}] \left( 2 \, \Gamma_{\beta \nu}^{\, \gamma} \bar{u}^{\left( \beta \right.} g_{\left. \gamma \right) \alpha} + \bar{u}^\beta \Gamma_{\beta \nu}^{\, \gamma} g_{\gamma \alpha} \right)
\end{eqnarray}
\begin{eqnarray}
    r_{\mu \alpha} (\tau; \bar{z}] &=& \frac{e^2}{3} \, g_{(2)}(r) \, w_{\mu \alpha} [\bar{z}]
\end{eqnarray}
\end{widetext}

\section{Noise-Induced Drift}

In this section we outline the details for obtaining the noise-induced force on the particle (\ref{drift}).  We begin with the ALD-Langevin equation with noise added in by hand
\begin{eqnarray}
    m_{ren} [z] \, \frac{ D u_\mu}{d\tau} = F_\mu ^{ext} (\tau) + f_\mu [z] + \eta_\mu ^+ [z]
\end{eqnarray}
If we expand the worldline coordinates in powers of the coupling constant $e$
\begin{eqnarray}
    z^\mu = z^\mu _0 + z^\mu _1 + z^\mu _2 + \ldots
\end{eqnarray}
where the subscript denotes the order of the expansion, then to lowest order one finds
\begin{eqnarray}
    m_{ren} \, \frac{D_0 u_{0\mu} }{d\tau} = F_\mu ^{ext} (\tau)
\end{eqnarray}
where $D_0/d\tau$ is the covariant $\tau$-derivative defined with respect to $z_0^\mu$. If the external force vanishes then this is nothing more than the geodesic equation.

The first order correction to the trajectory can be found by solving
\begin{eqnarray}
    && \frac{d u_{1\mu} }{d\tau} - 2 \Gamma_{\alpha \mu} ^{\, \beta} (z_0) u_0 ^{\left( \alpha \right.} u _{1\left. \beta \right)} - z_1 ^\gamma \partial_{z_0^\gamma} \Gamma_{\alpha \mu} ^{\, \beta} (z_0) u_0 ^\alpha u_{0\beta} \non \\
    && {\hskip0.5in} = \frac{ \eta_\mu ^+ (z_0) }{ m_{ren} }
\end{eqnarray}
which has the formal solution
\begin{eqnarray}
    z_1 ^\mu (\tau) = \int d\tau^\prime \, g_{ret} ^{\mu \nu} (\tau, \tau^\prime) \, \eta_\nu ^+ (z_0 ^{\alpha^\prime} )
\end{eqnarray}
and $g_{ret}^{\mu\nu}$ is the retarded Green's function for $z_1$.

To second order in $e$ we find that $z_2$ satisfies
\begin{widetext}
\begin{eqnarray}
    && m_{ren} \left( \frac{d u_{2\mu} }{d\tau} - 2 \Gamma_{\alpha \mu} ^{\, \beta} (z_0) u_0 ^{\left( \alpha \right.} u _{2\left. \beta \right)} - z_2 ^\gamma \partial_{z_0^\gamma} \Gamma_{\alpha \mu} ^{\, \beta} (z_0) u_0 ^\alpha u_{0\beta} \right) \nonumber \\
    &&{\hskip1.5in} = e^2 \frac{F_\mu ^{ext}}{m_{ren}} \int _{\tau_i}^\tau d\tau^\prime \, V( z_0 ^\alpha, z_0 ^{\alpha^\prime} )+  f_\mu [z_0] + z_1 ^\nu \nabla_{z_0^\nu} \eta_\mu ^+ (z_0) \nonumber \\
    &&{\hskip 1.75in} + m_{ren} \left( \Gamma_{\alpha \mu} ^{\, \beta} (z_0) u_1 ^\alpha u _{1\beta} + 2 z_1 ^\gamma \partial_{z_0^\gamma} \Gamma_{\alpha \mu} ^{\, \beta} (z_0)  u_1 ^{\left( \alpha \right.} u_{0\left. \beta \right)} + \frac{1}{2} z_1 ^\gamma z_1 ^\delta \partial _{z_0 ^\gamma} \partial _{z_0 ^\delta} \Gamma_{\alpha \mu} ^\beta (z_0)  u_0 ^\alpha u_{0\beta} \right) \non \\
\end{eqnarray}
Taking the stochastic average we find that the terms after the self-force give a non-vanishing contribution to the force on the particle
\begin{eqnarray}
    F_\mu ^{drift} (\tau) &=& \bigg\langle z_1 ^\nu \nabla_{z_0^\nu} \eta_\mu ^+ (z_0) + m_{ren} \left( \Gamma_{\alpha \mu} ^{\, \beta}  u_1 ^\alpha u _{1\beta} + 2 z_1 ^\gamma \partial_{z_0^\gamma} \Gamma_{\alpha \mu} ^{\, \beta} u_1 ^{\left( \alpha \right.} u_{0\left. \beta \right)}  + \frac{1}{2} z_1 ^\gamma z_1 ^\delta \partial _{z_0 ^\gamma} \partial _{z_0 ^\delta} \Gamma_{\alpha \mu} ^{\,\beta} u_0 ^\alpha u_{0\beta}  \right) \bigg\rangle _{\eta^+} \nonumber
\end{eqnarray}
Substituting in for $z_1$ we find that $F_\mu ^{drift}$ can be written in terms of the noise kernel associated with $\eta_\mu ^+$
\begin{eqnarray}
    F_\mu ^{drift}(\tau) &=& \int d\tau^\prime \, g_{ret} ^{\nu \sigma^\prime} \, \nabla_{z_0^\nu} \big\langle \eta_\mu ^+ (z_0 ^\lambda) \, \eta_{\sigma^\prime} ^+ (z_0 ^{\lambda^\prime}) \big\rangle _{\eta^+} \nonumber \\
    &+& m_{ren} \int d\tau^\prime \, d\tau^{\prime\prime} \left\{ \frac{1}{2} \,\partial _{z_0 ^\gamma} \partial _{z_0 ^\delta} \Gamma_{\alpha \mu} ^{\,\beta} (z_0^\lambda) u_{0\beta} \, g_{ret} ^{\gamma \rho^\prime} \, g_{ret} ^{\delta \sigma^{\prime\prime}}  + 2 \partial_{z_0^\gamma} \Gamma_{\alpha \mu} ^{\, \beta} (z_0^\lambda) \, g_{ret} ^{\gamma \rho^\prime} \, u_{0\left( \beta \right.} \frac{d}{d\tau} \, g_{ret}^{\left. \alpha \right) \sigma^{\prime\prime}} \right. \nonumber \\
    &&{\hskip1.1in} \left. + \Gamma _{\alpha \mu}^{\, \beta} (z_0 ^\lambda) \left( \frac{d}{d\tau} \, g_{ret}^{\alpha \rho ^\prime} \right)  \left( \frac{d}{d\tau} \, g_{ret \, \beta} ^{~~~~~\sigma^{\prime\prime}} \right) \right\} \big\langle \eta_{\rho^\prime} ^+ (z_0 ^{\lambda^\prime}) \, \eta_{\sigma^{\prime \prime}} ^+ (z_0 ^{\lambda^{\prime\prime}}) \big\rangle _{\eta^+}  \label{drift_force}
\end{eqnarray}
Notice that the first term in (\ref{drift_force}) comes from linearizing the stochastic force whereas the other terms have their origins in the kinetic term $m[z] a_\mu$. Finally, we arrive at (\ref{drift}) by defining the kernel $F_\mu ^{\rho \sigma}$ as the integrand appearing above
\begin{eqnarray}
    F_\mu ^{drift} (\tau) = \int d\tau^\prime \, d\tau^{\prime\prime} \, F_\mu ^{\rho\sigma} (\tau, \tau^\prime, \tau^{\prime\prime} ) \, \big\langle \eta_\rho ^+ (z_0 ^{\lambda^\prime}) \, \eta_\sigma ^+(z_0 ^{\lambda^{\prime\prime}}) \big\rangle _{\eta^+} \non \\
\end{eqnarray}
\end{widetext}

%\newpage


\begin{thebibliography}{99}
\bibitem{footnote_1} In general, for systems with a nonlinear coupling to the environment or for non-Gaussian initial states of the environment the interpretation of $P_\xi$ as a probability distribution is not always possible. It may take on negative values in which case $P_\xi$ should be interpreted as a pseudo-probability distribution in a similar vein as the Wigner function \cite{Calzetta_Roura_Verdaguer}.
\bibitem{footnote_2} Recently, the authors of \cite{Goldberger_Rothstein} have used the methods of effective field theory to describe the gravitational two-body problem of extended objects as an effective point-particle theory.
\end{thebibliography}
\end{document}